\documentclass[showkeys,amsmath,amssymb,pre]{revtex4}
\usepackage{color}
\usepackage{graphicx}
\usepackage{bbm}
\usepackage{amssymb,amsbsy,amsmath}
\usepackage{undertilde}
\newtheorem{lemma}{Lemma}

\newtheorem{ass}{Assumption}
\newtheorem{asse}{Assertion}

\newtheorem{theorem}{Theorem}

\newcommand{\re}{{\mathrm e}}
\newcommand{\ri}{{\mathrm i}}
\newcommand{\kB}{k_{\rm B}}

\begin{document}

\title[Periodic thermodynamics]
      {Periodic thermodynamics of a two spin Rabi model}

\author{Heinz-J\"urgen Schmidt}
	
\affiliation{Universit\"at Osnabr\"uck, Fachbereich Physik,
 	D-49069 Osnabr\"uck, Germany}

\begin{abstract}
We consider two $s=1/2$ spins with Heisenberg coupling and a monochromatic, circularly polarized magnetic field
acting only onto one of the two spins. This system turns out to be analytically solvable.
Also the statistical distribution of the work performed by the driving forces during one period can be
obtained in closed form and the Jarzynski equation can be checked.
The mean value of this work, viewed as a function of the physical parameters, exhibits features that can be related to some kind of Rabi oscillation.
Moreover, when coupled to a heat bath the two spin system will approach a non-equilibrium steady state
(NESS) that can be calculated in the golden rule approximation. The occupation probabilities of the NESS
are shown not to be of Boltzmann type, with the exception of a single phase with infinite quasitemperature.
The parameter space of the two spin Rabi model can be decomposed into
eight phase domains such that the NESS probabilities possess discontinuous derivatives at the phase boundaries.
The latter property is shown to hold also for more general periodically driven $N$-level systems.
\end{abstract}


\maketitle


\section{Introduction}
\label{sec:I}
A quantum system developing according to a time-dependent Hamiltonian $H(t)$
which varies periodically with time~$t$, such that
\begin{equation}
	H(t) = H(t+T) \; ,
\end{equation}	
possesses a complete set of {\em Floquet states\/}, that is, of solutions to
the time-dependent Schr\"odinger equation having the particular form
\begin{equation}
	\psi_n(t)= u_n(t)\, \exp(-\ri\varepsilon_n t) \; .
\label{eq:FST}
\end{equation}
The {\em Floquet functions\/} $ u_n(t) $ are also $T$-periodic
and
the quantities $\varepsilon_n$ are known as {\em quasienergies\/}~\cite{Zeldovich66,
Sambe73,FainshteinEtAl78}. They are only  uniquely determined up to integer
multiples of the driving frequency $\omega=\frac{2\pi}{T}$.

The significance of these Floquet states~(\ref{eq:FST}) is based on the fact that
every solution $ \psi(t)$ to the time-dependent Schr\"odinger equation
can be expanded with respect to the Floquet basis,
\begin{equation}
	 \psi(t) = \sum_n c_n \,  	
	u_n(t)\, \exp(-\ri\varepsilon_n t) \; ,
\end{equation}
such that the coefficients $c_n$ do not depend on time. Hence, the Floquet states
propagate with constant occupation probabilities~$| c_n |^2$, despite the
presence of a time-periodic drive. However, if the
periodically driven system is interacting with an environment, as it happens
in many cases of experimental interest~\cite{BlumelEtAl91,GrifoniHanggi98,
GasparinettiEtAl13,StaceEtAl13,ZhangEtAl17,ChoiEtAl17}, that environment may
continuously induce transitions among the system's Floquet states.
This has the effect that after some relaxation time
a quasi\-stationary distribution $\{ p_n \}$ of Floquet-state
occupation probabilities is reached which contains no memory of the
initial state. The question arises how to quantify this distribution.

In a short programmatic note entitled ``Periodic Thermodynamics'',
Kohn~\cite{Kohn01}
has drawn attention to such quasi\-stationary Floquet-state distributions
$\{ p_n \}$.  In an earlier
pioneering study, Breuer {\em et al.\/} had already calculated these
distributions for time-periodically forced oscillators coupled to a thermal
oscillator bath~\cite{BreuerEtAl00}.
To date, a great variety of different individual aspects of the ``periodic thermodynamics''
envisioned by Kohn has been discussed in the literature~\cite{{KetzmerickWustmann10},HoneEtAl09,
BulnesCuetaraEtAl15,ShiraiEtAl15,Liu15,IadecolaEtAl15a,IadecolaEtAl15,
SeetharamEtAl15,VorbergEtAl15,VajnaEtAl16,RestrepoEtAl16,LazaridesMoessner17,
SeetharamEtAl19}, but a coherent overall picture is still lacking.

In this situation it seems advisable to resort to models which are sufficiently
simple to admit analytical solutions. Recent results into this direction are the following:

\begin{itemize}
  \item As mentioned above, for the particular case of a linearly forced
harmonic oscillator the authors of \cite{BreuerEtAl00} have shown that the
Floquet-state distribution remains a Boltzmann distribution
with the temperature of the heat bath, see also \cite{LangemeyerHolthaus14}.
\item Similarly, the parametrically driven harmonic oscillator assumes a quasi-stationary state with a quasi-temperature
that is, however, generally different from the bath temperature, see \cite{DiermannEtAl19}, \cite{DiermannHolthaus19}.
  \item A spin $s$  exposed to both a static magnetic field and an oscillating,
circularly polarized magnetic field applied perpendicular
to the static one, as in the classic Rabi set-up \cite{Rabi37},
and coupled to a thermal bath of harmonic oscillators has been shown to approach a quasi Boltzmann distribution, see \cite{SSH19}.
This work generalizes the results of \cite{LangemeyerHolthaus14} for the case $s=1/2$.
\end{itemize}

\begin{figure}[t]
\centering
\includegraphics[width=0.7\linewidth]{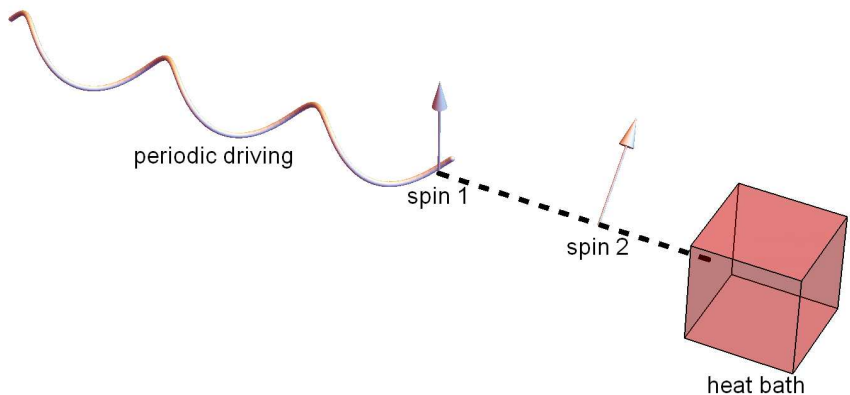}
\caption{Schematic representation of the two spin Rabi model considered in this paper.
}
\label{FIGPRIN}
\end{figure}

In the present work we will consider, similarly as in \cite{LangemeyerHolthaus14}, an $s=1/2$ spin with a circularly
polarized driving but only coupled to the heat bath via another $s=1/2$ spin, see Figure \ref{FIGPRIN}.
An analogous system has previously been numerically investigated with the focus on decoherence \cite{AJN06}.
In order to keep the analytical treatment as simple as possible we will set $\omega=\omega_0=1$, where $\omega_0$ denotes the
dimensionless Larmor frequency of the static magnetic field. Then it is possible to explicitly calculate
the quasienergies $\epsilon_n$  and the probabilities $p_n,\;n=1,\ldots,4$ of the NESS, although the latter are too complex
to be given in closed form. It turns out that the $p_n$ are {\em not\/} of Boltzmann type thereby rigorously confirming
the general conjectures about the nature of the NESS for a simple system. Another result will be the partition
of the parameter space $\boldsymbol{\mathcal P}$ into certain  {\em phases\/} $\boldsymbol{\mathcal P}_\nu$
such that the $p_n$, while being smooth functions of the parameters within the phases  $\boldsymbol{\mathcal P}_\nu$,
will have discontinuous derivatives at the phase boundaries. These findings will also hold for
general periodically driven $N$-level systems. For the special system under consideration we additionally observe that
all four NESS probabilities coincide for a certain phase $A$ which could be formally understood as an infinite
quasitemperature of this phase. But we will provide arguments that this result is confined to this very system
and will probably not hold in general.

The paper is organized as follows. In Section \ref{sec:D} we define the system to be studied and derive its explicit
time evolution in the Floquet normal form. The time evolution matrix for one period (monodromy matrix)
of the present system turns out to be symmetric and hence possesses real eigenvectors.
The proof of this has been moved to an Appendix \ref{sec:AP}. The explicit results on the time evolution are used in
Section \ref{sec:W} to calculate the statistical distribution of the work
performed by the periodic driving during one period and to check our results by confirming the corresponding Jarzynski equation
As a by-product we prove the physically  plausible fact that the expectation value of the work is always non-negative
and discuss the mean value of the work.
The general golden-rule approach to periodic thermodynamics is briefly recapitulated in Section \ref{sec:PGR}
and applied to the two spin system under consideration in Section \ref{sec:PA}. The partition of the parameter space
into phases and the $2^{nd}$ order phase transitions at the phase boundaries seems to hold also for the general
case of periodically driven $N$-level systems. The pertinent arguments are presented in the Appendix \ref{sec:AE}.
We close with a summary and outlook in Section  \ref{sec:SO}.

\section{Definitions and general results}
\label{sec:D}
We consider two spins with $s=1/2$ and the composite system described by the four-dimensional Hilbert space
${\mathcal H}={\mathbbm C}^2 \otimes {\mathbbm C}^2$.
The static Hamiltonian is assumed to be of the form
\begin{equation}\label{D1}
 H_0=\utilde{\mathbf s}^{(1)}_3\otimes {\mathbbm 1}+ {\mathbbm 1}\otimes \utilde{\mathbf s}^{(2)}_3+\lambda\; \utilde{\mathbf s}^{(1)}\cdot\utilde{\mathbf s}^{(2)}
 \;,
\end{equation}
where $\utilde{\mathbf s}^{(1)}$ and  $\utilde{\mathbf s}^{(2)}$ are the usual  $s=1/2$ vector spin operators for the subsystems
and $\lambda>0$ is some coupling parameter.
The eigenvalues $E_n$ of $H_0$ are
\begin{equation}\label{D1a}
 E_{1,2}=\frac{\lambda}{4}\pm 1,\;E_3=-\frac{3 \lambda }{4},\;E_4=\frac{\lambda }{4}
 \;.
\end{equation}

The periodic circularly polarized driving with amplitude $f$ and unit angular frequency acts only on the first spin and thus the total Hamiltonian
can be written as
\begin{equation}\label{D2}
  H(t)=H_0+f\,\left(\cos t \,\utilde{\mathbf s}^{(1)}_1+\sin t\,\utilde{\mathbf s}^{(1)}_2\right)
  \;.
\end{equation}
Upon choosing the eigenbasis of
$\utilde{\mathbf s}^{(1)}_3\otimes \utilde{\mathbf s}^{(2)}_3$ symbolically written as
$(\uparrow\uparrow,\uparrow\downarrow,\downarrow\uparrow,\downarrow\downarrow)$
this Hamiltonian can be identified with the  Hermitean $4\times 4$-matrix:
\begin{equation}\label{D3}
 H(t)=\left(
\begin{array}{cccc}
 \frac{\lambda +4}{4} & 0 & \frac{1}{2} f \re^{-i t} & 0 \\
 0 & -\frac{\lambda }{4} & \frac{\lambda }{2} & \frac{1}{2} f \re^{-i t} \\
 \frac{1}{2} f \re^{i t} & \frac{\lambda }{2} & -\frac{\lambda }{4} & 0 \\
 0 & \frac{1}{2} f \re^{i t} & 0 & \frac{\lambda -4}{4} \\
\end{array}
\right)
\;.
\end{equation}

First we will solve the corresponding Schr\"odinger equation ($\hbar=1$)
\begin{equation}\label{D4}
 \ri\,\frac{\partial}{\partial t}\psi(t)=H(t)\,\psi(t)
 \;.
 \end{equation}
 To this end we differentiate (\ref{D4}) three times w.~r.~t.~$t$ and eliminate all components
 of $\psi(t)$ except the first one $\psi_1(t)$. This yields a linear $4^{th}$ order differential equation
 for $\psi_1(t)$ of the form:
 \begin{eqnarray}\label{D5}\nonumber
 \frac{\partial^4}{\partial t^4}\psi_1(t)&=&
  -\frac{1}{256} (2 f-\lambda -4) (2 f+\lambda +4) \left(4 f^2+(\lambda +4) (3 \lambda-4)\right) \psi_1(t)\\
  \label{D5}
  && -\frac{\ri}{8} \left(
  \left(8 f^2+(\lambda -2) (\lambda +4)^2\right)  \frac{\partial}{\partial t}\psi_1(t)
  -\ri \left(4 f^2+3 \lambda ^2-48\right)
    \frac{\partial^2}{\partial t^2}\psi_1(t)
       +32 \frac{\partial^3}{\partial t^3}\psi_1(t)\right)   \;.
 \end{eqnarray}
 Remarkably, the coefficients of this differential equations are independent of $t$ due to the circularly polarized
 form of the driving. In contrast to the present case, for a linearly polarized driving of an
 $s=1/2$ spin the analogous elimination of the second  component of $\psi(t)$
 leads to a $2^{nd}$ order differential equation with $t$-dependent coefficients.
 Although this equation can be transformed into a confluent Heun equation, see \cite{MaLi07}, \cite{XieHai10}, and \cite{SSH20},
 it is by far more intricate than the $4^{th}$ order differential equation obtained in this paper.

 In our case the differential equation (\ref{D5}) can be elementarily solved by an exponential ansatz
 \begin{equation}\label{D6}
   \psi_1(t)=\sum_{n=1}^{4}c_n \,\exp\left(\ri\,\omega_n\,t \right)
   \;,
 \end{equation}
 with arbitrary coefficients $c_n\in{\mathbbm C}$.
 The $\omega_n$ can be obtained as the roots of an equation of $4^{th}$ order and assume the form:
 \begin{eqnarray}
 \label{D7a}
   \omega_1 &=&\frac{1}{4} \left(-2 \sqrt{f^2+\lambda ^2}+\lambda -4\right)\;, \\
   \label{D7b}
  \omega_2 &=& \frac{1}{4} \left(2 \sqrt{f^2+\lambda ^2}+\lambda -4\right)\;, \\
  \label{D7c}
   \omega_3 &=& \frac{1}{4} (-2 f-\lambda -4)\;, \\
   \label{D7d}
   \omega_4 &=& \frac{1}{4} (2 f-\lambda -4)\;.
 \end{eqnarray}
 If we would have included more parameters in the Hamiltonian (\ref{D2}), e.~g., the frequency $\omega$ of the periodic
 driving, this result would still be valid, albeit with a more complicated form of the roots that practically rules out a further
 analytical treatment of the problem.

The remaining three components of $\psi(t)$ are obtained by means of the following equations previously used for
eliminating $\psi_2(t),\psi_3(t),\psi_4(t)$:
 \begin{eqnarray}
\label{D8a}
   \psi_2(t) &=&-\frac{e^{\ri\, t}}{4 \,f\, \lambda }
   \left(16\left(2\,\ri\,\frac{\partial\psi_1}{\partial t}+ \frac{\partial^2\psi_1}{\partial t^2}\right)+\left(-16+4 f^2+\lambda ^2\right) \psi _1 \right)\;,  \\
   \label{D8b}
   \psi_3(t)&=& \frac{\ri\, e^{\ri \,t}}{2\, f}\left( 4\, \frac{\partial\psi_1}{\partial t} +\ri (\lambda +4)\psi _1\right)\;, \\
   \nonumber
   \psi_4(t) &=& \frac{e^{2\, \ri\, t}}{8 \,f^2  \,\lambda } \left(
   -4 \ri \left(\frac{\partial\psi_1}{\partial t} \left(4 f^2+5 \lambda ^2+8 \lambda -48\right)-4 \ri
   (\lambda -12)\frac{\partial^2\psi_1}{\partial t^2}+16 \frac{\partial^3\psi_1}{\partial t^3}\right.\right.\\
   \label{D8c}
   &&\left.\left.+ \left((\lambda +4)^2 (3 \lambda -4)-4 f^2 (\lambda -4)\right)\psi _1\right)\right)\;.
 \end{eqnarray}

Inserting $\psi_1(t)$ according to (\ref{D6}) and (\ref{D7a}-\ref{D7d}) into (\ref{D8a}-\ref{D8c}) yields
a first solution $\psi^{(1)}(t)$ that will be rewritten as
\begin{equation}\label{D9}
  \psi^{(1)}(t)=U(t)\;\left(\begin{array}{c}
                                 c_{1} \\
                                 c_{2} \\
                                 c_{3} \\
                                 c_{4}
                               \end{array} \right)\;,
\end{equation}
where $U(t)$ is a unitary $4\times 4$-matrix satisfying
\begin{equation}\label{D10}
 \frac{\partial}{\partial t}U(t)=-\ri\,H(t)\,U(t)
 \;.
\end{equation}
From this we obtain the fundamental system of solutions $\Psi(t)$ by
\begin{equation}\label{D11}
 \Psi(t)\equiv U(t)\,U(0)^{-1}
 \;,
\end{equation}
satisfying
\begin{equation}\label{D12}
 \Psi(0)={\mathbbm 1}
 \;.
\end{equation}
We will only explicitly give $\Psi(t)$ in its {\em Floquet normal form}
\begin{equation}\label{D13}
\Psi(t)={\mathcal P}(t)\,e^{-\ri {\mathcal F}\,t}
\;,
\end{equation}
such that ${\mathcal P}(t)$ is $2\pi$-periodic and ${\mathcal F}$ is the Floquet matrix.
After some calculations we obtain
\begin{equation}\label{D14}
 {\mathcal P}(t)=\left(
\begin{array}{cccc}
 \re^{-\ri\, t} & 0 & 0 & 0 \\
 0 & 1 & 0 & 0 \\
 0 & 0 & 1 & 0 \\
 0 & 0 & 0 & \re^{\ri\, t} \\
\end{array}
\right)
\;,
\end{equation}
and
\begin{equation}\label{D15}
 \re^{-\ri {\mathcal F}\,t} = A\,\Delta(t)\,A^\top
 \;,
\end{equation}
where
\begin{equation}\label{D16}
 A=\frac{1}{2}
 \left(
\begin{array}{cccc}
 -\sqrt{1-\alpha} & 1 & \sqrt{1+\alpha} & 1 \\
 -\sqrt{1+\alpha} & -1 & -\sqrt{1-\alpha} & 1 \\
 \sqrt{1+\alpha} & -1 & \sqrt{1-\alpha} & 1 \\
 \sqrt{1-\alpha} & 1 & -\sqrt{1+\alpha} & 1 \\
\end{array}
\right)
\;,
\end{equation}
setting
\begin{equation}\label{D16a}
  \alpha\equiv \frac{\lambda}{\sqrt{f^2+\lambda ^2}}
  \;,
\end{equation}
and
\begin{equation}\label{D17}
  \Delta(t)=\left(
\begin{array}{cccc}
 \re^{\frac{1}{4} \ri\, t \,\left(2 \sqrt{f^2+\lambda ^2}+\lambda \right)} & 0 & 0 & 0 \\
 0 & \re^{\frac{1}{4} \ri\, t\, (2 f-\lambda )} & 0 & 0 \\
 0 & 0 & \re^{\frac{1}{4} \ri\, t\, \left(\lambda -2 \sqrt{f^2+\lambda ^2}\right)} & 0 \\
 0 & 0 & 0 & \re^{-\frac{1}{4} \ri\, t\, (2 f+\lambda )} \\
\end{array}
\right)\;.
\end{equation}
The connection to the Floquet functions $u_n(t)$ mentioned in the Introduction is given by
\begin{equation}\label{D17a}
  u_n(t)={\mathcal P}(t)\, A_n
  \;,
\end{equation}
where $A_n$ denotes the $n$-th column of $A$.

We note the following special features of the form of (\ref{D13}) not yet fully understood.
First, it is not a priori clear that
according to (\ref{D14}) the periodic part ${\mathcal P}(t)$ is diagonal in the spin basis
and hence $[{\mathcal P}(t_1),{\mathcal P}(t_2)]=0$ for all $t_1,\,t_2\in{\mathbbm R}$.
Second, the eigenvectors of the Floquet matrix ${\mathcal F}$ that are the columns of $A$
according to (\ref{D16}) are real. This follows also from the fact the monodromy matrix $\Psi(2\pi)$
is unitary and symmetric, the latter property being a consequence of the particular structure of
the Hamiltonian (\ref{D3}), see Appendix \ref{sec:AP}. Note also
that the second and the fourth eigenvector is independent of $f$ and $\lambda$. 
These special properties of the monodromy matrix may explain the occurrence of the phase boundaries described in Section \ref{sec:PA}
despite the effect of ``avoided level crossing", see also the corresponding remarks in Section \ref{sec:AE}.

The quasienergies $\epsilon_n$ (eigenvalues of ${\mathcal F}$) can be directly read off the diagonal elements of (\ref{D17})
that represent the eigenvalues of $e^{-\ri {\mathcal F}\,t}$:
\begin{eqnarray}
\label{D18a}
  \epsilon_1 &=&  -\frac{1}{4}\left(2 \sqrt{f^2+\lambda ^2}+\lambda \right)\;, \\
  \label{D18b}
   \epsilon_2 &=& \frac{1}{4}  (\lambda-2f )\;. \\
   \label{D18c}
   \epsilon_3 &=& \frac{1}{4}\left(2 \sqrt{f^2+\lambda ^2}-\lambda \right)\\
   \label{D18d}
   \epsilon_4 &=& \frac{1}{4}  (\lambda+2f )\;.
\end{eqnarray}

Recall that the quasienergies are uniquely determined only up to integer multiples of $\omega=1$.
In (\ref{D18a}-\ref{D18d}) we have chosen representatives of quasienergies
that appear in a strictly monotonic increasing order for $\lambda,\,f>0$
which facilitates the calculations in the periodic thermodynamics section \ref{sec:PA}.
For the sake of consistency we will check the two limits $\lambda\to 0$ and $f\to 0$.

The static limit $f\to 0$ yields
\begin{equation}\label{D19}
 \lim_{f\to 0}\epsilon_2= \lim_{f\to 0}\epsilon_3= \lim_{f\to 0}\epsilon_4=\frac{\lambda}{4},\quad
 \mbox{and}\quad
 \lim_{f\to 0}\epsilon_1= -\frac{3\lambda}{4}
 \;.
\end{equation}
This agrees with the eigenvalues (\ref{D1a}) of the static Hamiltonian $H_0$ modulo integers.

The limit $\lambda\to 0$ means that the two spins are decoupled and hence the quasienergies should
approach those of the usual Rabi problem for the first spin plus the energy eigenvalues $\pm \frac{1}{2}$ of the second spin.
We obtain
 \begin{equation}\label{D20}
 \lim_{\lambda\to 0}\epsilon_3= \lim_{\lambda\to 0}\epsilon_4=\frac{f}{2},\quad
 \mbox{and}\quad
 \lim_{\lambda\to 0}\epsilon_1= \lim_{\lambda\to 0}\epsilon_2= -\frac{f}{2}
 \;.
\end{equation}
This has to be compatible with
\begin{equation}\label{D21}
 \epsilon_{\rm Rabi}=\frac{\omega\pm\Omega}{2}
 \;,
\end{equation}
where $\Omega$ is the Rabi frequency
\begin{equation}\label{D22}
 \Omega=\sqrt{f^2+(\omega_0-\omega)^2}
 \;.
\end{equation}
In our case we have chosen $\omega_0=\omega=1$ which implies $\Omega=f$ and further
$\epsilon_{Rabi}=\frac{1\pm f}{2}$. The total quasienergy of the decoupled spin system is thus
$\epsilon=\frac{1\pm f}{2}\pm\frac{1}{2}$.
Again, this is, modulo integers, in accordance with (\ref{D20}).

\section{Work performed on a two spin system}\label{sec:W}

As an application of the results obtained in the preceding Section \ref{sec:D} we consider the work performed on a two level system
by a circularly polarized magnetic field during one period.
In contrast to classical physics this work is not just a number but,
following \cite{TLH07}, has to be understood in terms of two subsequent energy measurements.
Before the time $t=0$ the two level system is assumed to be in a mixed state according to the canonical ensemble
\begin{equation}\label{W1}
 W=\exp\left(-\beta H_0\right)/\mbox{Tr}\left(\exp\left(-\beta H_0\right)\right)
 \;,
 \end{equation}
 with dimensionless inverse temperature $\beta=\frac{\hbar \,\omega}{\kB\,T}$ and $H_0$ being the static Hamiltonian (\ref{D1}).
Then  at the time $t=0$ one performs a L\"uders measurement of the instantaneous energy $H_0$ with the four possible
outcomes $E_n,\,n=1,\ldots,4$ according to (\ref{D1a}).
Hence after the measurement
the system is in the pure state $P_n$ with probability $\mbox{Tr}\left( P_n W\right)=\frac{1}{Z}\re^{-\beta E_n}, \;n=1,\ldots,4$, where
the $P_n$  are the projectors onto the eigenstates of $H_0$, {\em i.e.\/},
\begin{equation}\label{W2}
 P_1=\left(
\begin{array}{cccc}
 1 & 0 & 0 & 0 \\
 0 & 0 & 0 & 0 \\
 0 & 0 & 0 & 0 \\
 0 & 0 & 0 & 0 \\
\end{array}
\right)
,\quad
   P_2=\left(
\begin{array}{cccc}
 0 & 0 & 0 & 0 \\
 0 & 0 & 0 & 0 \\
 0 & 0 & 0 & 0 \\
 0 & 0 & 0 & 1 \\
\end{array}
\right)
,\quad
P_3=\frac{1}{2}
\left(
\begin{array}{cccc}
 0 & 0 & 0 & 0 \\
 0 & 1 & -1 & 0 \\
 0 & -1 & 1 & 0 \\
 0 & 0 & 0 & 0 \\
\end{array}
\right)
,\quad
P_4=\frac{1}{2}
\left(
\begin{array}{cccc}
 0 & 0 & 0 & 0 \\
 0 & 1 & 1 & 0 \\
 0 & 1 & 1 & 0 \\
 0 & 0 & 0 & 0 \\
\end{array}
\right)
,
\end{equation}
and $Z=\sum_{n=1}^{4}\re^{-\beta\,E_n}$.
After this measurement the system evolves according to the Schr\"odinger equation (\ref{D4}) with Hamiltonian $H(t)$.
At the time $t=2\pi$ the system hence is in the pure state $\Psi(2\pi)\,P_n\, \Psi(2\pi)^\ast$ with probability $\mbox{Tr}\left( P_n W\right)$
for $n=1,\ldots,4$. Then a second measurement
of the static energy $H_0$ is performed, again with the four possible outcomes $E_n$. Both measurements together have
$4\times 4=16$ possible outcomes symbolized by pairs $(i,j)$ where $i,j=1,\ldots 4$ that occur with probabilities
\begin{equation}\label{W3}
 p(i,j)=\mbox{Tr}\left(W\,P_i\right) \mbox{Tr}\left(P_j\,\Psi(2\pi)\,P_i\,\Psi(2\pi)^\ast\right)
 \;,
\end{equation}
such that $\sum_{i,j=1}^{4}p(i,j)=1$.
We will not display the $p(i,j)$ but rather the marginal probabilities $p(i)\equiv\sum_{j=1}^{4}p(i,j)$ and the conditional
probabilities $\pi(j|i)\equiv\frac{p(i,j)}{p(i)}$, the latter being independent of $\beta$.
It is plausible and can be directly verified that the matrix of conditional
probabilities will be symmetric and hence doubly stochastic, see \cite{SG19} for the r\^ole of
double stochasticity in connection with the Jarzynski equation.
Thus we need only to display the values of
$\pi(j|i)$ for $j\le i$. The detailed results are
\begin{equation}\label{W4}
p(1)=\frac{1}{z}e^{2\beta},\quad p(2)=\frac{1}{z}e^{\beta(1+\lambda)},\quad p(3)=\frac{1}{z}e^{\beta},\quad p(4)
=\frac{1}{z}\equiv\frac{1}{e^{\beta } \left(e^{\beta  \lambda }+e^{\beta }+1\right)+1},
\end{equation}
and
\begin{eqnarray}
\label{W5a}
  \pi(1|1) &=& \pi(4|4)=a+b,\\
  \label{W5b}
  \pi(1|4)&=& a-b,\\
  a&=& \frac{1}{8} \left(\frac{f^2 \cos \left(2 \pi  \sqrt{f^2+\lambda ^2}\right)+2 f^2+3
   \lambda ^2}{f^2+\lambda ^2}+\cos (2 \pi  f)\right),\\
   \label{W5c}
   b&=& \frac{1}{2} \cos (\pi  f) \left(\frac{\lambda  \sin (\pi  \lambda ) \sin \left(\pi
   \sqrt{f^2+\lambda ^2}\right)}{\sqrt{f^2+\lambda ^2}}+\cos (\pi  \lambda ) \cos
   \left(\pi  \sqrt{f^2+\lambda ^2}\right)\right),\\
   \label{W5d}
  \pi(1|2) &=& \pi(2|4)=\frac{f^2 \sin ^2\left(\pi  \sqrt{f^2+\lambda ^2}\right)}{2 \left(f^2+\lambda
   ^2\right)} ,\\
   \label{W5e}
  \pi(2|2) &=& \frac{f^2 \cos \left(2 \pi  \sqrt{f^2+\lambda ^2}\right)+f^2+2 \lambda ^2}{2
   \left(f^2+\lambda ^2\right)} ,\\
   \label{W5f}
  \pi(2|3) &=& 0
  \;.
\end{eqnarray}
Besides the symmetry of the matrix of conditional probabilities there are additional coincidences in (\ref{W5a}), (\ref{W5d})
and vanishing values in (\ref{W5f}) that are not yet understood.

The matrix of probabilities $p(i,j)$ contains all information for the probability distribution of the energy differences between the first and
the second measurement, {\em i.e.\/}, of the distribution of the work $w$ performed on the two spin system by means of the periodic driving.
Interestingly, although ``work" cannot be considered as an observable in the ordinary sense  giving rise to a projection-valued
measure \cite{TLH07}, it is an observable in the generalized sense of a positive-operator-valued measure \cite{RCP14}, \cite{BLPY16}.

\begin{figure}[t]
\centering
\includegraphics[width=0.7\linewidth]{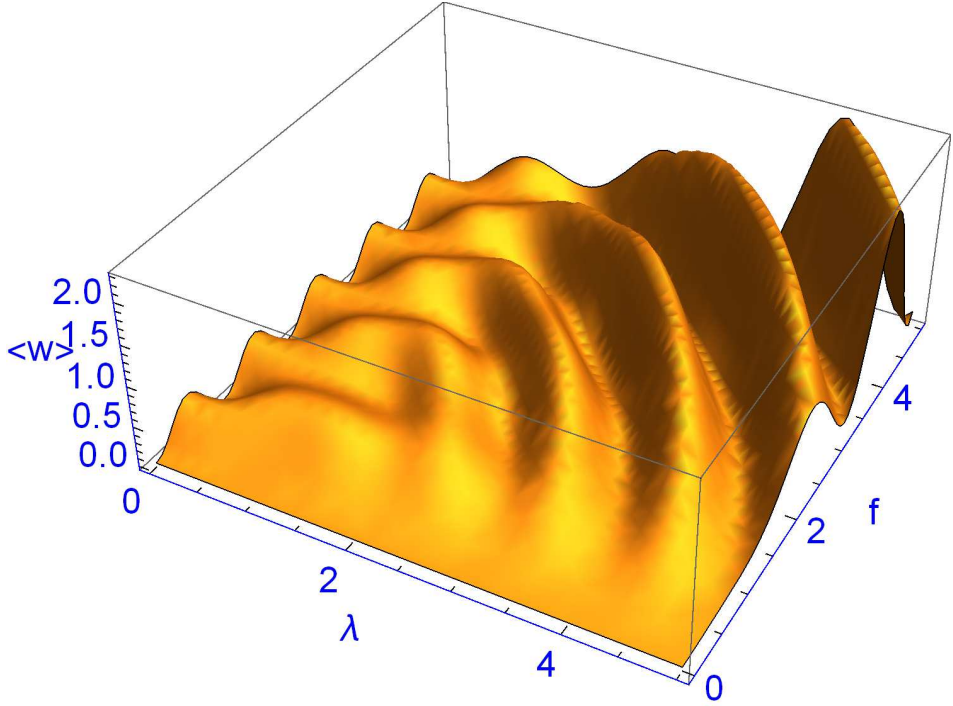}
\caption{The mean value $\langle w \rangle$ of the work performed on the two spin Rabi system
during one period as a function of the physical parameters $\lambda$ and $f$, where the initial inverse
temperature of the system has been set to $\beta=1$.
}
\label{FIGPW1}
\end{figure}

\begin{figure}[t]
\centering
\includegraphics[width=0.7\linewidth]{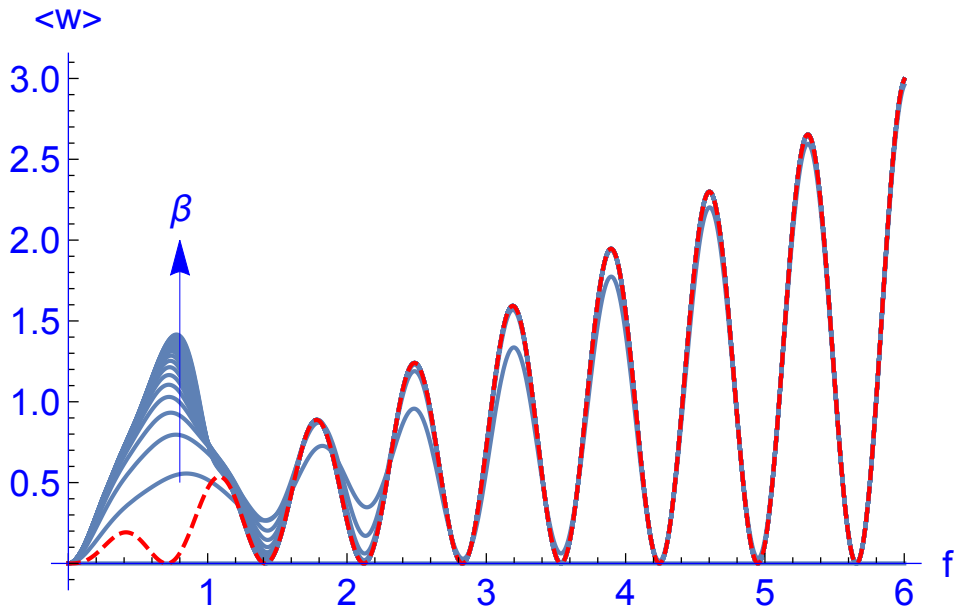}
\caption{The mean value $\langle w \rangle$ of the work performed on the two spin Rabi system
during one period as a function of the physical parameters $\lambda=f$ and $\beta=0,1,\ldots, 20$, where the
increasing values of $\beta$ are indicated by an arrow. Moreover, we show the asymptotic form of
$\langle w \rangle\sim \frac{1}{2} f \sin ^2\left(\sqrt{2}\, \pi \, f\right)$ (red, dashed curve).
}
\label{FIGPW2}
\end{figure}

For example, we may calculate the mean value of the performed work with the result
\begin{eqnarray} \label{W6}
  \left\langle w\right\rangle &=&\sum_{i,j=1}^{4}(E_j-E_i)\,p(i,j)=\frac{1}{4 (f^2 + \lambda^2)z}\left(w_1+w_2+w_3\right)\;,\\
  \label{W6a}
  w_1 &=&4 \left(e^{2 \beta }-1\right) \lambda ^2-f^2 \left(e^{2 \beta } (\lambda -4)-2
   \lambda  e^{\beta  \lambda +\beta }+\lambda +4\right)\;,\\
   \label{W6b}
  w_2 &=&f^2 \lambda  \left(-2 e^{\beta  \lambda +\beta }+e^{2 \beta }+1\right) \cos \left(2
   \pi  \sqrt{f^2+\lambda ^2}\right)-8 e^{\beta } \sinh (\beta ) \left(f^2+\lambda^2\right)
   \cos (\pi  f) \cos (\pi  \lambda ) \cos \left(\pi  \sqrt{f^2+\lambda^2}\right)\\
   \label{W6c}
   w_3&=&-4 \left(e^{2 \beta }-1\right) \lambda  \sqrt{f^2+\lambda ^2} \cos (\pi  f) \sin (\pi
    \lambda ) \sin \left(\pi  \sqrt{f^2+\lambda ^2}\right)
  \;,
\end{eqnarray}
where the parameter $z$ in (\ref{W6}) has been defined in (\ref{W4}). This function is shown in Figure \ref{FIGPW1}
for the inverse temperature $\beta=1$. First, we note that obviously  $\langle w \rangle\ge 0$ which appears physically
plausible and will be proven below.

Another conspicuous feature of the graph of $\langle w \rangle(\lambda,f,1)$ is its oscillating behaviour
with increasing amplitude for large values of $\lambda\approx f$.
This will be more clearly demonstrated in the Figure \ref{FIGPW2} where we have set $\lambda=f$
and displayed $\langle w \rangle(f,f,\beta)$ for values of $\beta=0,1,\ldots,20$. It is obvious from this Figure and can be
analytically confirmed that
\begin{equation}\label{W7a}
\langle w \rangle(f,f,\beta) \sim \frac{1}{2} f \sin ^2\left(\sqrt{2}\, \pi \, f\right)
\mbox{ for } f\to\infty
\;.
\end{equation}
The convergence of $\langle w \rangle(f,f,\beta)$ against its asymptotic behaviour holds for all $\beta\ge 0$ but will be
more rapid for large $\beta$. We will give a semi-quantitative explanation. For large $\beta$, i.~e., low temperatures
the system is practically in its ground state with energy $E_3$ at $t=0$, the begin of the periodic driving, see (\ref{D1a}).
By the driving it will be excited to the next lowest state with energy $E_2$. The probability of excitation
$p_{3\rightarrow 2}(t)$ can be calculated and yields a rather simple expression for the special case $\lambda=f$:
\begin{equation}\label{W7b}
 p_{3\rightarrow 2}(t)=\frac{1}{4} \sin ^2\left(\frac{f\, t}{\sqrt{2}}\right)
 \;.
\end{equation}
This result is analogous to the well-known Rabi oscillation of a two-level system. It is further plausible
that the mean value of the work during one period will be maximal if some maximum of (\ref{W7b}) will
be attained after exactly one period of driving, i.~e., at $t=2\pi$. This happens for
\begin{equation}\label{W7c}
 \frac{f\, 2\,\pi}{\sqrt{2}}=\frac{n\,\pi}{2},\quad n\;\mbox{being odd}\quad \Leftrightarrow f=\frac{n}{2 \sqrt{2}}
 \;,
\end{equation}
and hence at the maxima of the asymptotic form of
$\langle w \rangle(f,f,\beta) \sim \frac{1}{2} f \sin ^2\left(\sqrt{2}\, \pi \, f\right)$.
An analogous reasoning applies to the minima of $\langle w \rangle(f,f,\beta)$
Hence the oscillating structure of $\langle w \rangle$ visible in the Figure \ref{FIGPW1} can be viewed
as a footprint of a kind of approximate  Rabi oscillation occurring for the two spin Rabi model.
Moreover, it is also plausible that asymptotically $\langle w \rangle(f,f,\beta)$ scales with $f$.\\

Finally, we may, after some calculations, confirm the famous Jarzynski equation \cite{TLH07} that in our case reads
\begin{equation}\label{W7}
 \left\langle e^{-\beta\,w}\right\rangle=\sum_{i,j=1}^{4}e^{-\beta\,(E_j-E_i)}\,p(i,j)= 1
  \;.
\end{equation}
The latter can be considered as a test of consistency of our results. Further, we may apply
Jensen's inequality to the convex function $x\mapsto -\log x$ and conclude
\begin{equation}\label{W8}
  \left\langle \beta\,w\right\rangle= \left\langle -\log\left(e^{-\beta\,w} \right) \right\rangle
  \stackrel{\rm Jensen}{\ge} -\log  \left\langle e^{-\beta\,w} \right\rangle
  \stackrel{(\ref{W7})}{=}-\log 1=0
  \;,
\end{equation}
which, due to $\beta>0$, means that the expectation value of the performed work is always non-negative which would be difficult to be confirmed
directly for the expression (\ref{W6}-\ref{W6c}) of $\left\langle w \right\rangle$.

\section{Periodic thermodynamics}
\label{sec:P}

\subsection{Golden-rule approach to open driven systems}
\label{sec:PGR}

Let us consider a quantum system evolving according to a $T=\frac{2\pi}{\omega}$-periodic
Hamiltonian~$H(t)$ on a Hilbert space ${\mathcal H}_S$  that is additionally
coupled to a heat bath, described by a
Hamiltonian $H_{\rm bath}$ acting on a Hilbert space ${\mathcal H}_{B}$.
The total Hamiltonian on the composite Hilbert space
${\mathcal H}_S \otimes {\mathcal H}_B$ takes the form
\begin{equation}\label{eq:HTOT}
	H_{\rm total}(t) = H(t) \otimes {\mathbbm 1}
	+ {\mathbbm 1} \otimes H_{\rm bath} + V\otimes W \; .
\end{equation}
Moreover, following Breuer {\em et al.\/}~\cite{BreuerEtAl00}, let us
consider a bath consisting of thermally occupied harmonic oscillators,
and an interaction of the  prototypical form
\begin{equation}
	W = \sum_{\widetilde\omega} \left(
	b_{\widetilde\omega}^{\phantom\dagger} + b_{\widetilde\omega}^\dagger
	\right) \; ,	
\label{eq:PTF}
\end{equation}
where $b_{\widetilde\omega}^{\phantom\dagger}$ ($b_{\widetilde\omega}^\dagger$)
is the annihilation (creation) operator pertaining to a bath oscillator of
frequency~$\widetilde{\omega}$.

For weak coupling the effect of the heat bath can be approximately described
by a variant of the Golden Rule. Since this approach has been elaborately explained
in the literature, see \cite{LangemeyerHolthaus14} and \cite{SSH19}, we will
confined ourselves here with the enumeration of the pertinent formulas sticking closely to \cite{SSH19}.

In the golden-rule approximation the heat bath induces transitions between the system's
Floquet states $u_i(t)$ and $u_f(t)$ with transition rates
$\Gamma_{fi}$ that can be written as sums over partial rates
\begin{equation}
	\Gamma_{fi} = \sum_{\ell\in{\mathbbm Z}} \, \Gamma_{fi}^{(\ell)} \; .
\label{eq:TOR}
\end{equation}
given by
\begin{equation}
	\Gamma_{fi}^{(\ell)} = 2\pi \, | V_{fi}^{(\ell)} |^2 \,
	N(\omega_{fi}^{(\ell)})  \, J(|\omega_{fi}^{(\ell)}|) \; .
\label{eq:GFI}
\end{equation}
Here $J(|\omega_{fi}^{(\ell)}|)$ denotes the spectral density of the frequency of bath phonons
and will be set to a constant $J_0>0$ in what follows. Further,
$V_{fi}^{(\ell)}$ denotes the Fourier components of the $T$-periodic matrix elements
\begin{equation}
	\tilde{V}_{fi}=\langle u_f(t) | \, V \, | u_i(t) \rangle
	= \sum_{\ell\in{\mathbbm Z}} \, V_{fi}^{(\ell)}
	\exp(\ri\ell\omega t) \; ,
\label{eq:FDV}	
\end{equation}
and $N(\omega_{fi}^{(\ell)})$ is the value of the function $N(\tilde{\omega})$
evaluated at
\begin{equation}\label{eq:DOM}
 \omega_{fi}^{(\ell)}\equiv\epsilon_f-\epsilon_i+\ell\,\omega
 \;.
\end{equation}
Physically, $N(\tilde{\omega})$ represents the thermal average of the
bath phonon occupation density and is given by
\begin{equation}\label{eq:DN}
 N(\tilde{\omega})=\left\{
 \begin{array}{r@{\quad:\quad}l}
 \tilde{\omega}>0 & \frac{1}{\exp(\beta\tilde{\omega})-1},\\
 \tilde{\omega}<0 & \frac{1}{1-\exp(\beta\tilde{\omega})},
 \end{array}
 \right.
\end{equation}
where $\beta$ is the inverse temperature of the bath,
not to be confounded with the inverse temperature considered in Section \ref{sec:W}.
The case distinction in (\ref{eq:DN}) corresponds to the distinction between the creation
of a bath phonon ($\tilde{\omega}>0$) and its absorption  ($\tilde{\omega}<0$).
Thus, a transition among Floquet states is not simply associated with only
one single frequency, but rather with a set of frequencies spaced by integer
multiples of the driving frequency~$\omega$, reflecting the ladder-like nature
of the system's quasienergies.

The total rates~(\ref{eq:TOR}) now determine the desired quasi\-stationary
distribution~$\{ p_n \}$ as a solution to the Pauli master equation~\cite{BreuerEtAl00}
\begin{equation}
	\sum_m \big( \Gamma_{nm} p_m - \Gamma_{mn} p_n \big) = 0 \; ,
\label{eq:PME}
\end{equation}	
where the existence of a strictly positive solution will be shown below.
According to this equation~(\ref{eq:PME}), the quasi\-stationary
distribution $\{ p_m \}$ which establishes itself under the
combined influence of time-periodic driving and the thermal oscillator bath
is the eigenvector of a matrix $\widetilde{\Gamma}$ corresponding
to the eigenvalue $0$, where $\widetilde{\Gamma}$ is obtained from
$\Gamma$ by subtracting from the diagonal
elements the respective column sums, {\em i.e.\/},
\begin{equation}
	\widetilde{\Gamma}_{mn} \equiv \Gamma_{mn}
	- \delta_{mn} \sum_{k=1}^{N}\Gamma_{kn} \; .
\label{eq:GAT}
\end{equation}	
Moreover,
it is evident that we only need the non-diagonal matrix elements of~$\Gamma$
for calculating the quasistationary distribution,
whereas the diagonal elements would be required for computing the dissipation
rate~\cite{LangemeyerHolthaus14}.

As announced above, we will now prove the existence of a strictly positive
solution of the Pauli master equation (\ref{eq:PME}). Although this
result it well-known it is not easily found in the literature and hence an
explicit proof will be in order.

We start with a few definitions needed for the statement of the theorem of Frobenius-Perron that is suited for the problem at hand.
A real $N\times N$-matrix $T$ will be called {\em non-negative}, in symbols $T\ge 0$,  iff all its matrix entries satisfy $T_{ij}\ge 0$.
Analogously, we will define a {\em positive} matrix $T>0$ and also use these terms for vectors $x$ with the notation $x>0$ or $x\ge 0$.
Moreover, $T$ is {\em irreducible} iff for all $1\le i,j\le N$ there exists a $k\in{\mathbbm N}$ such that
$T^k_{ij}>0$. Physically, if $T$ is some transition matrix, the notion of irreducibility would be construed as a kind of ``ergodicity",
because it says that if starting from any state $i$ it is possible to reach any other state $j$ after a finite number of steps.
Then we may state the theorem of Frobenius-Perron, see, e.~g., \cite{G59}, Theorem 2, p.~53, in the following form, adapted to our purposes.
\begin{theorem}\label{TFP}(Frobenius-Perron)\\
Let $T$ be a non-negative irreducible square matrix. Then
\begin{itemize}
  \item $T$ has a positive eigenvalue $\lambda_{\rm max}$ that is the {\em spectral radius} of $T$, i.~e., all other
    eigenvalues $\lambda$ of $T$ satisfy $\left|\lambda\right|\le \lambda_{\rm max}$.
  \item Furthermore $\lambda_{\rm max}$ has algebraic and geometric multiplicity
    one, and has an eigenvector $x$ with $x > 0$.
  \item Any non-negative eigenvector of $T$ is a multiple of $x$.
\end{itemize}
\end{theorem}

By means of (\ref{eq:GFI}) it is obvious that $\Gamma\ge 0$, but the present two spin Rabi model is
an example showing that $\Gamma> 0$ does not hold in general, see below.
Hence, in order to apply the preceding theorem, we will additionally need the following
\begin{ass}\label{AS1}
  $\Gamma$ is irreducible,
\end{ass}
that is essentially saying that the eigenvectors of the interaction matrix $V$ are oblique w.~r.~t.~the Floquet basis
and does not follow from the general assumptions made so far.

Recall that $\widetilde{\Gamma}$  is defined by subtraction of the column sums of $\Gamma$ and hence
will possess negative matrix entries in the diagonal. If  $\lambda$ is defined as the maximal
column sum of $\Gamma$ we will obtain a non-negative matrix $G$ by adding $\lambda$ to each diagonal element,
\begin{equation}\label{eq:G}
  G\equiv \widetilde{\Gamma}+\lambda {\mathbbm 1}\ge 0
  \;,
\end{equation}
and, moreover, conclude
\begin{lemma}\label{LG}
  $G$ and hence also $G^\top$ are irreducible.
\end{lemma}

{\bf Proof}: By definition, $G$ can be written as $G=\Gamma+\Delta$ such that $\Delta\ge 0$ is a diagonal matrix.
It follows from
\begin{equation}\label{eq:Bin}
 G^k =\left( \Gamma+\Delta\right)^k= \Gamma^k+\Delta\,\Gamma^{k-1}+ \ldots + \Gamma^{k-1}\Delta+\ldots +\Delta^2\Gamma^{k-2}+
 \ldots+ \Gamma^{k-2}\Delta^2+\ldots+\Delta^k
 \;,
\end{equation}
and the Assumption \ref{AS1} that for all $1\le i,j\le N$ there exists a $k\in{\mathbbm N}$ such that $G_{ij}^k>0$. Hence $G$ is irreducible.
 \hfill$\Box$\\
 By definition, $\widetilde{\Gamma}$ has vanishing column sums, hence ${\mathbf 1}\equiv (1,1,\ldots,1)$ will be a left
 eigenvector of $\widetilde{\Gamma}$ with eigenvalue $0$. It follows that ${\mathbf 1}$ is also a right eigenvector
 of $G^\top$ with eigenvalue $\lambda$. $G^\top$ satisfies the conditions of the theorem of Frobenius-Perron, hence
 $\lambda=\lambda_{\rm max}$ is the spectral radius of $G^\top$ and ${\mathbf 1}$ is the unique corresponding eigenvector.
 Applying again the theorem of Frobenius-Perron to $G$ that has the same eigenvalues as $G^\top$ we conclude that
 there exists an eigenvector $p>0$ of $G$ with eigenvalue $\lambda$, unique up to normalization. It follows
 that $\widetilde{\Gamma}\,p=0$ and hence $p$ is the solution of the Pauli master equation (\ref{eq:PME}) we are seeking for.
We state this result as
\begin{theorem}\label{TUP}
  If the matrix $\Gamma$ is irreducible then the Pauli master equation (\ref{eq:PME}) has a unique solution
  $\{p_n\}$ satisfying   $p_n>0$ for all $n=1,\ldots N$ and $\sum_{n=1}^{N}p_n=1$.
\end{theorem}

\subsection{Application to the two spin system}\label{sec:PA}

\begin{figure}[t]
\centering
\includegraphics[width=0.7\linewidth]{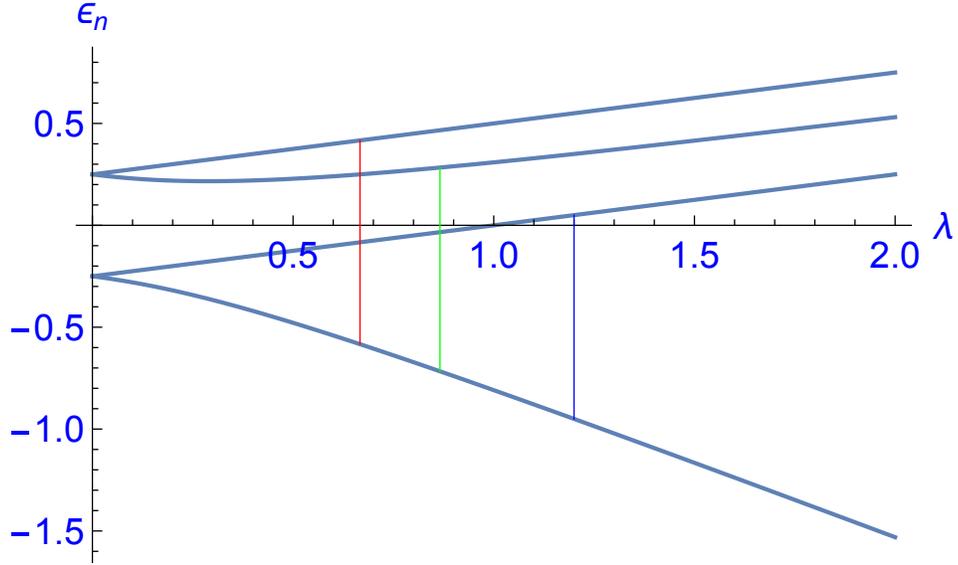}
\caption{The four quasienergies $\epsilon_n$ according to (\ref{D18a}-\ref{D18d}) as functions of $\lambda$
where $f$ has been set to $1/2$. At the values
of $\lambda=\frac{2}{3},\frac{\sqrt{3}}{2}$ and $\frac{6}{5}$ certain differences of quasienergies
assume the value $1$ and hence the corresponding frequencies $\omega_{fi}^{(\ell)}$ according to (\ref{eq:DOM}) vanish.
These cases are indicated by vertical coloured lines. They correspond to certain phase boundaries in Figure \ref{FIGB}.
}
\label{FIGA1}
\end{figure}

\begin{figure}[t]
\centering
\includegraphics[width=0.7\linewidth]{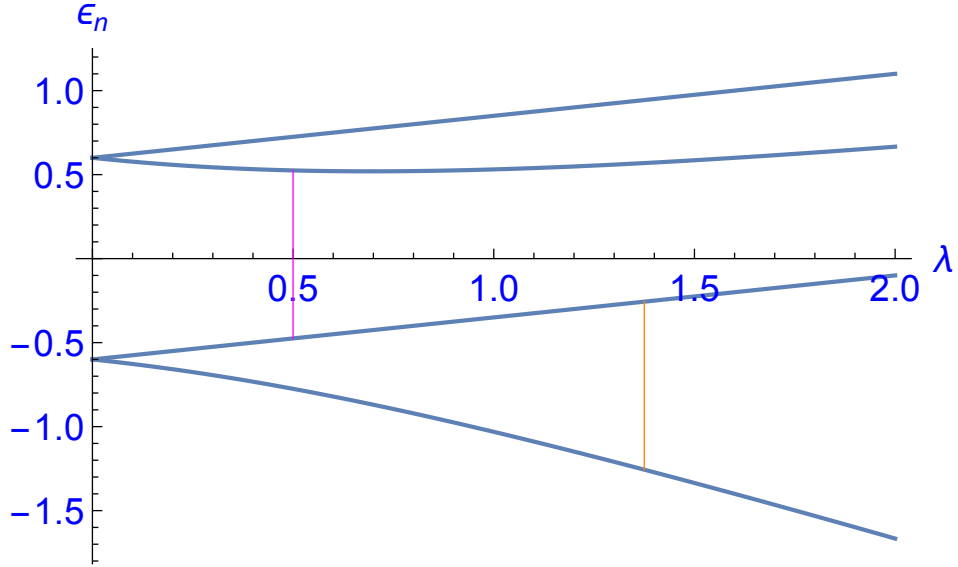}
\caption{Analogous to Figure \ref{FIGA1} but with $f=\frac{6}{5}$. Here the frequencies $\omega_{fi}^{(\ell)}$  vanish
at $\lambda=\frac{1}{2}$ and $\lambda=\frac{11}{8}$.
}
\label{FIGA2}
\end{figure}

\begin{figure}[t]
\centering
\includegraphics[width=0.7\linewidth]{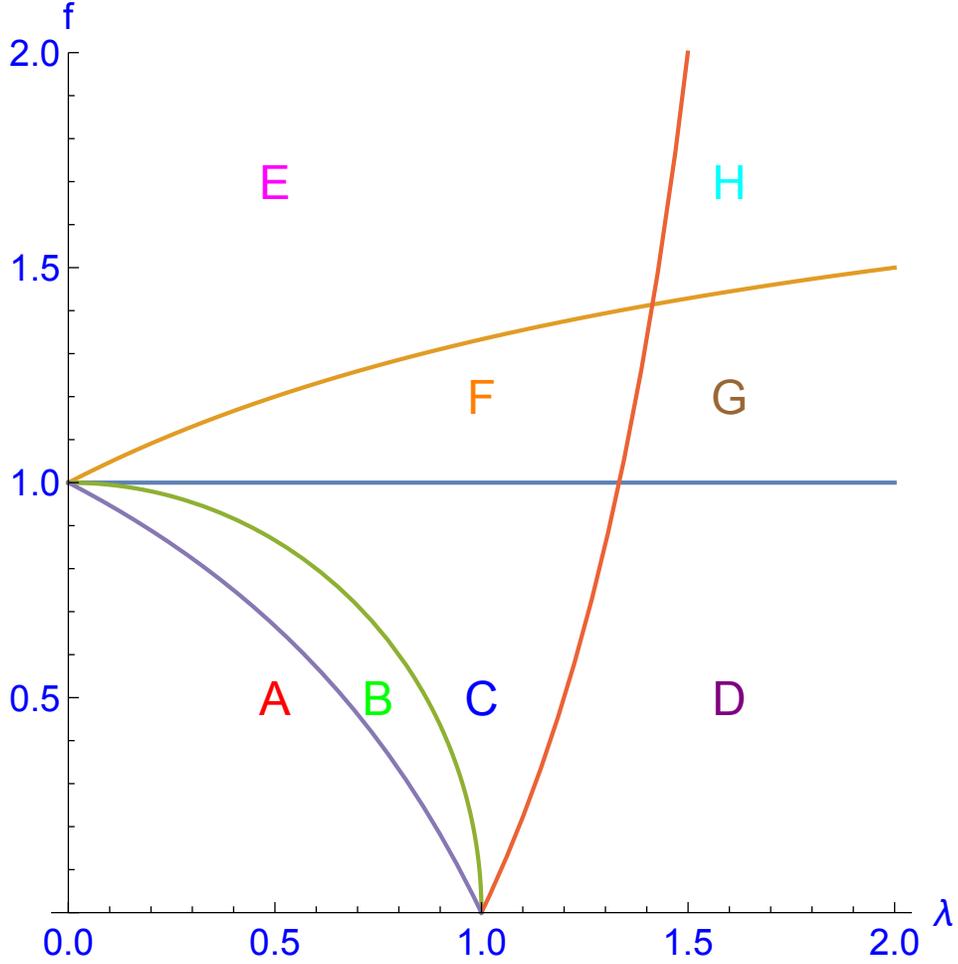}
\caption{The phase diagram of the $(\lambda,f)$-parameter space with eight phases $A,\ldots H$,
where the phase boundaries are given by the equations
(\ref{PA6}) (green circle) or (\ref{PA7a}-\ref{PA7d}) (red, yellow, purple, blue curves).
}
\label{FIGB}
\end{figure}

We choose the matrix $V$ that is part of the coupling to the heat bath according to (\ref{eq:HTOT}) as
$V\equiv{\mathbbm 1}\otimes {\mathbf s}^{(2)}_1$, {\em i.e.\/}, only the second spin is involved.
We need its matrix elements
$\tilde{V}_{fi}\equiv \langle u_f(t) | \, V \, | u_i(t) \rangle$
w.~r.~t.~Floquet states, see (\ref{eq:FDV}). In our case $\tilde{V}$ can be written as
\begin{equation}\label{PA1}
 \tilde{V}=A^\ast\, {\mathcal P }(t)^\ast\, V\,{\mathcal P }(t)\,A
 \;,
\end{equation}
with ${\mathcal P}(t)$ and $A$ according to (\ref{D14}) and (\ref{D16}).
It is clear from (\ref{D14}) that $\tilde{V}$ contains only Fourier components
of the order $|\ell|\le 1$. Actually, we obtain
\begin{equation}\label{PA2}
  \tilde{V}= V^{(1)}\,e^{\ri t}+V^{(-1)}\,e^{-\ri t}
  \;,
\end{equation}
where
\begin{equation}\label{PA3}
 V^{(1)}=\frac{1}{8}\left(  \begin{array}{cccc}
 2 f u & v+w & -2 \lambda  u & v-w \\
 -v-w & -2 & v-w & 0 \\
 -2 \lambda  u & w-v & -2 f u & v+w \\
 w-v & 0 & -v-w & 2 \\
\end{array}
\right)
\;,
\end{equation}

\begin{equation}\label{PA4}
 V^{(-1)}=\frac{1}{8}
\left(
\begin{array}{cccc}
 2 f u & -v-w & -2 \lambda  u & w-v \\
 v+w & -2 & w-v & 0 \\
 -2 \lambda  u & v-w & -2 f u & -v-w \\
 v-w & 0 & v+w & 2 \\
\end{array}
\right)
\;,
\end{equation}
and
\begin{equation}\label{PA5}
 u\equiv \frac{1}{\sqrt{f^2+\lambda ^2}},\quad v\equiv\sqrt{1+\lambda  u},\quad w\equiv \sqrt{1-\lambda  u}
 \;.
\end{equation}
Note that the occurrence of the matrix entry $0$ in (\ref{PA3}) and (\ref{PA4}) implies that $\Gamma_{24}=\Gamma_{42}=0$
and hence $\Gamma$ is not positive but only non-negative which has to be taken into account in the application
of Theorem \ref{TFP}.

Further we need the values of $N(\omega_{fi}^{(\ell)})$ in (\ref{eq:GFI}) according to (\ref{eq:DN}).
Recall that the case distinction to be made w.~r.~t.~the sign of
$\omega_{fi}^{(\ell)}=\epsilon_f-\epsilon_i+\ell\omega=\epsilon_f-\epsilon_i+\ell$
physically corresponds to the absorption or generation of bath phonons.
In order to obtain analytical expressions for, say, the occupation probabilities in the
non-equilibrium steady state (NESS), we will have to restrict the parameters $(\lambda, f)\in{\mathbbm R}_+\times {\mathbbm R}_+$
to certain domains where the sign of $\omega_{fi}^{(\ell)}$ will not change for all $f,i,\ell$.
These domains can be viewed as ``phases" of a phase diagram of the parameter space ${\mathbbm R}_+\times {\mathbbm R}_+$.
The boundaries of these phases are given by equations of the form $\omega_{fi}^{(\ell)}=0$. The latter corresponds to
a partial degeneracy of quasienergies taking into account that they are only defined up to integer multiples
of the driving frequency $\omega=1$.

We consider the example $f=3,\,i=1,$ and $\ell=-1$. The corresponding boundary equation is
\begin{equation}\label{PA6}
0=\omega_{31}^{(-1)}= \epsilon_3-\epsilon_1-1=\frac{1}{4} \left(-\lambda +2 \sqrt{f^2+\lambda ^2}\right)+\frac{1}{4} \left(\lambda
   +2 \sqrt{f^2+\lambda ^2}\right)-1=\sqrt{f^2+\lambda ^2}-1
\;,
\end{equation}
describing a quarter circle in the $(\lambda,f)$-quadrant, see Figure \ref{FIGB}.

The other boundaries are given by
\begin{eqnarray}
\label{PA7a}
  0=\epsilon_2-\epsilon_1-1 &\Leftrightarrow& f=\frac{2 (\lambda -1)}{2-\lambda }\;, \\
  \label{PA7b}
 0=\epsilon_3-\epsilon_2-1 &\Leftrightarrow& f=\frac{2 (\lambda +1)}{2+\lambda }\;, \\
 \label{PA7c}
 0=\epsilon_4-\epsilon_1-1 &\Leftrightarrow&f=\frac{2 (\lambda -1)}{\lambda -2} \;,\\
 \label{PA7d}
  0=\epsilon_4-\epsilon_2-1 &\Leftrightarrow& f=1\;
\end{eqnarray}
see the Figures \ref{FIGA1}, \ref{FIGA2} and \ref{FIGB}. Note that there are six positive differences of
quasienergies $\epsilon_f-\epsilon_i$ but only five boundary equations since the equation
$\epsilon_4-\epsilon_3-1=0$ has no positive solution.

As a first, somewhat surprising analytical result we note that for the phase $A$ defined by $f<\frac{2 (\lambda -1)}{\lambda -2}$, see
Figure \ref{FIGB}, the Pauli master equation (\ref{eq:PME}) has a unique solution corresponding to the same
occupation probability for all Floquet states.
This also follows from the symmetry $\Gamma_{mn}=\Gamma_{nm}$ that holds only within phase $A$.
Formally the coincidence of all probabilities would correspond to an infinite quasitemperature
and could be compared with the vanishing inverse quasitemperature along the line $\omega=\omega_0$ and $0<F<\omega_0$
for the circularly polarized Rabi problem, see \cite{SSH19}, figure $1$.

\begin{figure}[t]
\centering
\includegraphics[width=0.7\linewidth]{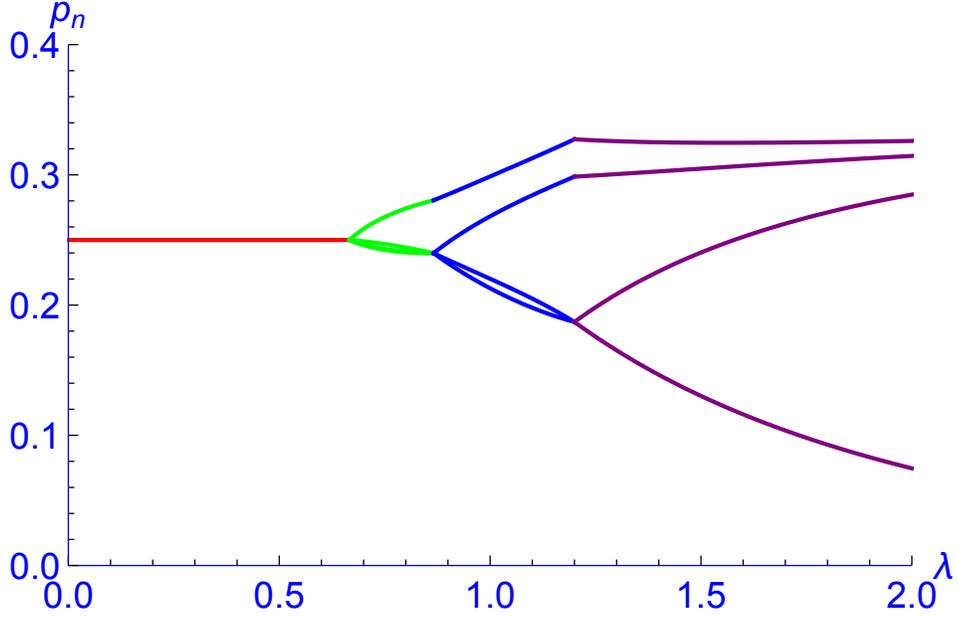}
\caption{The four occupation probabilities $p_n$ of the Floquet states for the NESS as functions of $\lambda$
where $f$ has been set to $f=1/2$ and the inverse bath temperature is chosen as $\beta=1$.
Within the phases $A$ -- $D$, indicated by different colours, the $p_n$ are smooth functions of $\lambda$.
At the phase boundaries the derivatives $\frac{d\,p_n}{d\,\lambda}$ are discontinuous and at least two probabilities coincide.
}
\label{FIGC}
\end{figure}

\begin{figure}[t]
\centering
\includegraphics[width=0.7\linewidth]{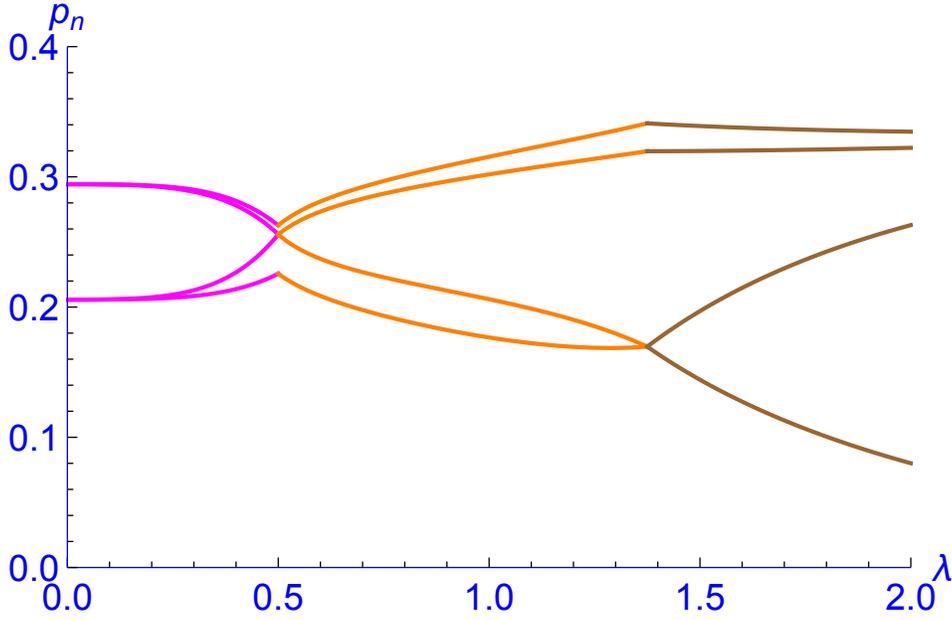}
\caption{The four occupation probabilities $p_n$ of the Floquet states for the NESS as functions of $\lambda$
where $f$ has been set to $f=6/5$ and $\beta=1$. Within the phases $E$, $F$, $G$, indicated by different colours,
the $p_n$ are smooth functions of $\lambda$.
At the phase boundaries the derivatives $\frac{d\,p_n}{d\,\lambda}$ are discontinuous and exactly two probabilities coincide.
}
\label{FIGD}
\end{figure}

In the phase domains $B$ -- $H$ the occupation probabilities $p_n$ can be analytically calculated by the means
of computer-algebraic software but the results cannot be displayed due to their forbidding complexity.
Nevertheless, one may plot these results. A first graphics shows the $p_n$ as continuous functions of $\lambda$ where
the parameter $f$ has been set to $f=1/2$, see Figure \ref{FIGC}. One clearly distinguishes the four phases
$A$ --$D$ acoording to Figure \ref{FIGB} and observes that the $p_n(\lambda)$ are smooth inside the phase domains
but shows kinks at the phase boundaries. The fact that at least two probabilities coincide at the phase boundaries
can be understood by the arguments presented in Appendix \ref{sec:AE} that also hold for general $N$-level systems.

The coincidence of two probabilities at phase boundaries also shows that, in general, the NESS will not be
of Boltzmann type with a quasitemperature $\theta$: For a Boltzmann distribution of occupation probabilities
$p_n$ and non-degenerate representatives of quasienergies two probabilities never coincide except for
$\theta=\infty$. In our case the latter only occurs in the phase $A$, see above.

\section{Summary and outlook}\label{sec:SO}
We have investigated the two spin Rabi model consisting of an $s=1/2$ spin subjected to a monochromatic
circularly polarized magnetic field and coupled to a second spin $s=1/2$ that is in turn in contact
with a heat bath. The quasienergies of the spin system as well as the occupation probabilities of the
emerging non-equilibrium steady state (NESS) can be, in principle, analytically determined and hence this system
may serve as an example for testing conjectures about general periodically driven $N$-level systems.
We found that, in contrast to other systems recently studied,
the NESS probabilities are not of Boltzmann type and hence there does not exist a quasitemperature.
Moreover, the parameter space of the system is found to be partitioned into certain phases such that the NESS
probabilities change at the phase boundaries in a way analogous to a $2^{nd}$ order phase transition.
It has been made plausible by detailed arguments that these two properties will also be satisfied
for general $N$-level systems. On the other hand, the existence of a phase $A$ with infinite quasitemperature
hinges on special properties of the two spin Rabi model, e.~g., the structure of the
eigenvectors of the Floquet operator or the commuting operators describing the periodic part of the time evolution,
and probably does not generally hold.
Nevertheless, it would be instructive to closer investigate similar systems in order to verify (or falsify) the above conjectures.

\appendix

\section{Proof of the symmetry of the monodromy matrix}
\label{sec:AP}
As noted in Section \ref{sec:D} the symmetry of the unitary monodromy matrix $U(2\pi)$ has
the consequence that it possesses a real eigenbasis. In fact, the eigenvalue equation
\begin{equation}\label{AP1}
 U(2\pi)\,\phi=c\,\phi
 \;,
\end{equation}
satisfying $|c|^2=1$ implies
\begin{equation}\label{AP1a}
\overline{\phi}=U(2\pi)\,U(2\pi)^{-1}\,\overline{\phi}=U(2\pi)\,\overline{U(2\pi)}\,\overline{\phi}
\stackrel{(\ref{AP1})}{=}\bar{c}\,U(2\pi)\,\overline{\phi}
\;,
\end{equation}
where we have used that, according to the above symmetry assumption, $\overline{U(2\pi)}=U(2\pi)^{-1}$.
This means that the vector $\overline{\phi}$ will be an eigenvector of $U(2\pi)$ corresponding
to the same eigenvalue $\frac{1}{\bar{c}}=c$. Thus if $\phi$ is unique it must be real, or otherwise,
in the case of degeneracy, it can be chosen as real.\\

It remains to show that $U(2\pi)$ is symmetric. To this end we introduce a slightly more general notation
by writing the unitary time evolution between $t=t_0$ and $t=t_1$ as $U(t_1,t_0)$ such that
\begin{equation}\label{AP1b}
 U(t_1,t_0)=U(t_0,t_1)^{-1}
 \;.
\end{equation}
$U(t,0)$  satisfies the differential equation
\begin{equation}\label{AP2}
  \frac{\partial}{\partial t}U(t,0)=-\ri\,H(t)\,U(t,0)
  \;,
\end{equation}
analogous to (\ref{D10}) and the initial condition $U(0,0)={\mathbbm 1}$.
Moreover,
\begin{equation}\label{AP3}
 U(t-2\pi,-2\pi)=U(t,0)
 \;,
\end{equation}
due to the $2\pi$-periodicity of $H(t)$.

Note that the
special form of the Hamiltonian (\ref{D3}) due to circular polarization of the driving field implies
\begin{equation}\label{AP4}
 \overline{H(t)}=H(-t)
 \;.
\end{equation}
Define the family of unitaries $V(t,0)\equiv \overline{U(-t,0)}$. It satisfies
\begin{equation}\label{AP5}
  \frac{\partial}{\partial t}V(t,0)=-\overline{\frac{\partial}{\partial t}U(-t,0)}
  \stackrel{(\ref{AP2})}{=}-\overline{\left(-\ri\, H(-t) U(-t,0)\right)} \stackrel{(\ref{AP4})}{=}-\ri\,H(t)\,V(t,0)
  \;,
\end{equation}
and $V(0,0)={\mathbbm 1}$, the same differential equation and initial condition as $U(t,0)$. Hence
\begin{equation}\label{AP6}
V(t,0)=U(t,0)=\overline{U(-t,0)}\quad\mbox{for all } t\in{\mathbbm R}
\;.
\end{equation}
Especially, for $t=2\,\pi$,
\begin{equation}\label{AP7}
U(2\pi,0)=\overline{U(-2\pi,0)}\stackrel{(\ref{AP1b})}{=}\overline{U(0,-2\pi)^{-1}}
\stackrel{(\ref{AP3})}{=}\overline{U(2\pi,0)^{-1}}=U(2\pi,0)^\top
\;,
\end{equation}
which completes the proof of $U(2\pi,0)$ being symmetric. \hfill$\Box$\\

\section{Some properties of periodically driven $N$-level systems}
\label{sec:AE}
We adopt a more general framework than in the main part of the paper
and assume a Hamiltonian $H(\boldsymbol{\pi},t)$ as an Hermitean $N\times N$-matrix
depending on certain parameters $\boldsymbol{\pi}\in\boldsymbol{\mathcal P}\subset{\mathbbm R}^p$
including the driving frequency $\omega$. Here the parameter space $\boldsymbol{\mathcal P}$
is assumed to be an open subset of ${\mathbbm R}^p$.
Again, the Hamiltonian
will depend $T\equiv\frac{2\pi}{\omega}$-periodically on $t$.
Moreover, we will assume that there exists a strictly monotone selection of quasienergies
$\epsilon_n(\boldsymbol{\pi}),\; n=1,\ldots,N$ that depend smoothly on $\boldsymbol{\pi}\in\boldsymbol{\mathcal P}$:
\begin{ass}\label{AS2}
 \begin{equation}\label{AE2}
 \epsilon_n(\boldsymbol{\pi})<\epsilon_m(\boldsymbol{\pi}) \mbox{ for all } 1\le n < m \le N \mbox{ and } \boldsymbol{\pi}\in\boldsymbol{\mathcal P}
 \;.
\end{equation}
\end{ass}

Analogously to the definitions in Section \ref{sec:PA} we will define ``phases" $\boldsymbol{\mathcal P}_{\nu}\subset\boldsymbol{\mathcal P}$
by intersections of open subsets of $\boldsymbol{\mathcal P}$ of the form
\begin{equation}\label{AE3}
{\mathcal O}_{nm\ell}^>\equiv\left\{\boldsymbol{\pi}\in\boldsymbol{\mathcal P}\left|  \epsilon_n(\boldsymbol{\pi})- \epsilon_m(\boldsymbol{\pi})+ \ell \omega>0  \right.\right\}
\end{equation}
or
\begin{equation}\label{AE4}
{\mathcal O}_{nm\ell}^<\equiv\left\{\boldsymbol{\pi}\in\boldsymbol{\mathcal P}\left|  \epsilon_n(\boldsymbol{\pi})- \epsilon_m(\boldsymbol{\pi})+ \ell \omega<0  \right.\right\}
\;.
\end{equation}
We are looking for ``minimal phases" in the sense that $\boldsymbol{\mathcal P}_{\nu}$ must not contain strictly smaller phases.
Although the integer $\ell$ in (\ref{AE3}) and (\ref{AE4}) may assume infinitely many values it suffices to consider
{\em finitely} many intersections of the above subsets. This can be seen as follows. Let $n>m$ such
$\epsilon_n(\boldsymbol{\pi})-\epsilon_m(\boldsymbol{\pi})>0$. Then there exists an $\ell\in{\mathbbm N}_0$
such that $\epsilon_n(\boldsymbol{\pi})-\epsilon_m(\boldsymbol{\pi})-\ell\,\omega>0$
but $\epsilon_n(\boldsymbol{\pi})-\epsilon_m(\boldsymbol{\pi})-(\ell+1)\,\omega<0$.
It follows that for the pair $(n,m)$ we need only consider the intersection of the two subsets ${\mathcal O}_{n,m,-\ell}^>$ and
${\mathcal O}_{n,m,-(\ell+1)}^<$ since the other ones of the form (\ref{AE3}) or (\ref{AE4}) are always larger
and hence not minimal. Analogous considerations apply for the case $n<m$.
It follows that the $\boldsymbol{\mathcal P}_{\nu}$ are open as finite intersections of open subsets of $\boldsymbol{\mathcal P}$.

The phase boundaries are again given by equations of the form
\begin{equation}\label{AE5}
 \epsilon_n(\boldsymbol{\pi})- \epsilon_m(\boldsymbol{\pi})+ \ell \omega=0
 \;,
\end{equation}
and will be denoted by $\boldsymbol{\mathcal P}_{nm\ell}$.
It may happen, as in the case of the two spin Rabi model, that not all phase boundaries given by equations of the form (\ref{AE5})
are realized since only a finite number of non-vanishing Fourier components of the relevant quantities exists.

Another problem is the requirement that the phase boundaries should have codimension one in  $\boldsymbol{\mathcal P}$
whereas the ``avoided level crossing" of quasienergies, see, e.~g., \cite{H16}, is an indication of a larger codimension.
To explain this problem in more detail we reconsider the $N\times N$ monodromy matrix $U(T,0)$ describing the unitary time evolution
of the system after one period $T$ and recall that the eigenvalues of $U(T,0)$ are in $1:1$ relation with equivalence
classes of quasienergies modulo $\omega$. A general unitary $N\times N$-matrix depends on $N^2$ real parameters, but the submanifold
of unitary matrices with one pair of degenerate eigenvalues has only the dimension $N^2 -3$, i.~e., the codimension three.
This supports the expectation that in the $p$-dimensional surface $\boldsymbol{\mathcal P}$ the phase boundaries given by (\ref{AE5})
should also have codimension three, and not one as required in our approach. 
Note, however, that for special cases like the class of symmetric unitary matrices, see Appendix \ref{sec:AP}, the codimension reduces to two.
Moreover, two eigenvalues of $U(T,0)$ belonging to different eigenvalues of a symmetry will not show the avoided level crossing,
see, e.~g., \cite{H16}. Another way to circumvent the above problem results when one of the parameters is the 
frequency of excitation $\omega$. This frequency is constant for the monodromy matrix and the sketched argument for codimension three does not apply. 
As an illustration we remark that for the one spin $s=1/2$ Rabi problem with quasienergy
$\epsilon_\pm=\frac{1}{2}\left( \omega\pm \Omega_{\rm Rabi}\right)$, see (\ref{D21}), the crossing of quasienergies
$\epsilon_+=\epsilon_- +\omega$ occurs for $\omega=\frac{f^2+\omega _0^2}{2 \omega _0}$. The latter indicates a 
codimension one of the phase boundary in spite of the noncrossing rule.

The general definitions of Section \ref{sec:PGR} also apply for the $N$ level case. We note the following
\begin{lemma}\label{L1}
\begin{equation}\label{AE6}
  V_{nm}^{(\ell)}=\overline{ V_{mn}^{(-\ell)}} \mbox{ for all } n,m =1,\ldots,N \mbox{ and } \ell\in{\mathbbm Z}
  \;.
\end{equation}
\end{lemma}

\noindent {\bf Proof} :
Recall that, due to $V$ being Hermitean,
\begin{equation}\label{AE7}
 \tilde{V}_{nm}\stackrel{(\ref{eq:FDV})}{=}\langle u_n(t) | \, V \, | u_m(t) \rangle
 = \sum_{\ell\in{\mathbbm Z}} V_{nm}^{(\ell)}\,\re^{\ri \,\ell\,\omega\,t}
 =\overline{\langle u_m(t) | \, V \, | u_n(t) \rangle}=\overline{\tilde{V}_{mn}}=
 \sum_{\ell\in{\mathbbm Z}} \overline{V_{mn}^{(\ell)}}\,\re^{-\ri \,\ell\,\omega\,t}
 =\sum_{\ell\in{\mathbbm Z}} \overline{V_{mn}^{(-\ell)}}\,\re^{\ri \,\ell\,\omega\,t}
 \;.
\end{equation}
The comparison of the coefficients of the first and the last Fourier series in (\ref{AE7}) yields the result. \hfill$\Box$\\

Next we will formulate some arguments in favour of the following Assertion, albeit not in a mathematically rigorous manner.
\begin{asse}\label{A1}
At least two NESS probabilities coincide at the phase boundaries.
\end{asse}

Consider a fixed boundary $\boldsymbol{\mathcal P}_{\bar{n}\bar{m}\bar{\ell}}$
that is defined by the vanishing of some frequency $\omega_{\bar{m}\bar{n}}^{(\bar{\ell})}$.
It follows from
\begin{equation}\label{AE8}
\omega_{\bar{m}\bar{n}}^{(\bar{\ell})}=\epsilon_{\bar{m}}-\epsilon_{\bar{n}}+\bar{\ell}\omega=
-\left( \epsilon_{\bar{n}}-\epsilon_{\bar{m}}-\bar{\ell}\omega\right)=-\omega_{\bar{n}\bar{m}}^{(-\bar{\ell})}
\;,
\end{equation}
see (\ref{eq:DOM}), that the complementary frequency $\omega_{\bar{n}\bar{m}}^{(-\bar{\ell})}$ vanishes too.
For these values the thermal averages $N(\omega_{\bar{m}\bar{n}}^{(\bar{\ell})})$ and $N(\omega_{\bar{n}\bar{m}}^{(-\bar{\ell})})$
diverge due to (\ref{eq:DN}). Hence close to the boundary these averages and the corresponding
transition rates $\Gamma_{\bar{m}\bar{n}}$ and $\Gamma_{\bar{n}\bar{m}}$ will assume arbitrary large values.
If the Pauli master equation (\ref{eq:PME}) is written in the form
\begin{equation}\label{PA8}
  	\sum_m \Gamma_{nm} p_m =\sum_m \Gamma_{mn} p_n   \;,
\end{equation}
it is obvious that for $n=\bar{n}$ both sides of (\ref{PA8}) are dominated by a single term
where $m=\bar{m}$ and hence
\begin{equation}\label{PA9}
  \Gamma_{\bar{n}\bar{m}} p_{\bar{m}} \approx \Gamma_{\bar{m}\bar{n}} p_{\bar{n}}
  \;.
\end{equation}
This approximation is to be understood in the sense that although both sides
of (\ref{PA9}) become arbitrarily large its difference remains bounded. This means that close to the
phase boundary we obtain a kind of ``local detailed balance" for the pair $(\bar{m},\bar{n})$.
On the other hand the matrix entries $ \Gamma_{\bar{n}\bar{m}}$ will be almost symmetric, {\em i.e.\/},
satisfy $\Gamma_{\bar{n}\bar{m}}\approx \Gamma_{\bar{m}\bar{n}}$ close to the
phase boundary. This can be shown as follows. Using
\begin{equation}\label{PA11}
 \left| V_{\bar{m}\bar{n}}^{(\bar{\ell})}\right|^2 =  \left| V_{\bar{n}\bar{m}}^{(-\bar{\ell})}\right|^2
 \;,
\end{equation}
see Lemma \ref{L1} in this Appendix,
the limit relation
\begin{equation}\label{PA12}
 \lim_{\tilde{\omega}\downarrow 0}\frac{N(\tilde{\omega})}{N(-\tilde{\omega})}=
  \lim_{\tilde{\omega} \downarrow 0}\frac{1-\re^{-\beta\tilde{\omega}}}{\re^{\beta\tilde{\omega}}-1}=
  \lim_{\tilde{\omega}\downarrow 0}\re^{-\beta\,\tilde{\omega}}=1
  \;,
\end{equation}
and (\ref{AE8}), we conclude
\begin{equation}\label{PA13}
 \Gamma_{\bar{m}\bar{n}}\approx  \Gamma_{\bar{m}\bar{n}}^{(\bar\ell)}
\stackrel{(\ref{eq:GFI})}{=}2\,\pi \,\left| V_{ \bar{m}\bar{n}}^{(\bar\ell)}\right|^2 N(\omega_{\bar{m}\bar{n}}^{(\bar\ell)})\,J_0
 \approx
 2\,\pi \,\left| V_{ \bar{n}\bar{m}}^{(-\bar\ell)}\right|^2 N(\omega_{\bar{n}\bar{m}}^{(-\bar\ell)})\,J_0
 =\Gamma_{\bar{n}\bar{m}}^{(-\bar\ell)}
 \approx
  \Gamma_{\bar{n}\bar{m}}
  \;.
\end{equation}

Consequently, when approaching the phase boundary, symbolically denoted by $\lim_{\tilde{\omega}\downarrow 0} $, we have
\begin{equation}\label{PA14}
\lim_{\tilde{\omega}\downarrow 0} \frac{p_{\bar{m}}}{p_{\bar{n}}}\stackrel{(\ref{PA9})}{=}
\lim_{\tilde{\omega}\downarrow 0} \frac{ \Gamma_{\bar{m}\bar{n}}}{ \Gamma_{\bar{n}\bar{m}}}\stackrel{(\ref{PA13})}{=} 1
 \;,
\end{equation}
which completes the arguments in favour of Assertion \ref{A1}. \hfill$\Box$\\

In the case of a single spin $s$ all quasienergy levels are equidistant, see eqs.~(53) and (54) in \cite{SSH19},
and thus the coincidence of two probabilities at the phase boundary implies that all probabilities $p_n$ are the same
and hence the inverse quasitemperature vanishes, see  \cite{SSH19}.

In the general case arguments analogous to those at the end of Section \ref{sec:PA} show that the NESS will not be
of Boltzmann type at least at the phase boundaries and, by continuity, in a small neighbourhood of the phase
boundaries. This supports the conjecture that the existence of a quasitemperature of the NESS is restricted
to very special systems.

Next we will address the question how the NESS probabilities $p_n$ are connected at the phase boundaries and
formulate the following
\begin{asse}\label{A2}
The NESS probabilities are continuous at the phase boundaries but their gradients are discontinuous there.
\end{asse}
We will provide some arguments in favour of this assertion that could probably be strengthen to a more rigorous proof.
To this end we consider a fixed phase boundary $\boldsymbol{\mathcal P}_{\bar{n}\bar{m}\bar{\ell}}$
given by the equation
\begin{equation}\label{PA15}
  0=\omega_{\bar{n}\bar{m}}^{(\bar{\ell})}=\epsilon_{\bar{n}}-\epsilon_{\bar{m}}+\bar{\ell}\,\omega
  \;,
\end{equation}
and will calculate the $p_n$ in a small neighbourhood of some point
$\boldsymbol{\pi}\in \boldsymbol{\mathcal P}_{\bar{n}\bar{m}\bar{\ell}}$.
We consider a curve through $\boldsymbol{\pi}$ perpendicular to $\boldsymbol{\mathcal P}_{\bar{n}\bar{m}\bar{\ell}}$
parametrized by the parameter
\begin{equation}\label{PA16}
 x\equiv \beta\,\omega_{\bar{n}\bar{m}}^{(\bar{\ell})}
 \;,
\end{equation}
such that $-\delta<x<\delta$ for some $\delta>0$
and $x=0$ corresponds to the point $\boldsymbol{\pi}\in \boldsymbol{\mathcal P}_{\bar{n}\bar{m}\bar{\ell}}$.

First we only consider the ``positive neighbourhood"  $\boldsymbol{\mathcal P}_{\bar{n}\bar{m}\bar{\ell}}^>$
 of $\boldsymbol{\mathcal P}_{\bar{n}\bar{m}\bar{\ell}}$
given by $\omega_{\bar{n}\bar{m}}^{(\bar{\ell})} > 0$ (such that also $x>0$) and restricted
in such a way that no other phase boundaries intersect $\boldsymbol{\mathcal P}_{\bar{n}\bar{m}\bar{\ell}}^>$.
We assume that a Taylor series representation
of $p_n$ holds in $\boldsymbol{\mathcal P}_{\bar{n}\bar{m}\bar{\ell}}^>$ with the first terms being of the form
\begin{equation}\label{PA17}
  p_n=p_{n0}+x\,p_{n1}+O(x^2)
  \;.
\end{equation}
We denote by $\Gamma^>$ and $\widetilde{\Gamma}^>$ the transition rate matrix functions \label{PA27e}
(\ref{eq:TOR}) and (\ref{eq:GAT}) restricted to the positive neighbourhood
$\boldsymbol{\mathcal P}_{\bar{n}\bar{m}\bar{\ell}}^>$. According to what has been said
the matrix entries $\Gamma_{nm}^>$ will be smooth functions of $x$  for $-\delta<x<\delta$
except for  $\Gamma_{\bar{n}\bar{m}}^>$ and
 $\Gamma_{\bar{m}\bar{n}}^>$ where the transition rates diverge for $x\to 0$.
Hence it is sensible to adopt Laurent series representations for the $\Gamma_{nm}^>$ that
are Taylor series for most cases but start with
an $\frac{1}{x}$-term in the latter two cases.

In particular, isolating the diverging terms, we may write
\begin{equation}\label{PA18}
 \widetilde{\Gamma}_{\bar{n}\bar{m}}^> ={\Gamma}_{\bar{n}\bar{m}}^>=
 2\pi J_0\left\{
 \left| V_{\bar{n}\bar{m}}^{(\bar{\ell})}\right|^2\frac{1}{\re^x-1}+
 \sum_{\stackrel{\ell\in{\mathbbm Z}}{\ell\neq\bar{\ell}}} \left| V_{\bar{n}\bar{m}}^{(\ell)}\right|^2
 N\left(\omega_{\bar{n}\bar{m}}^{\ell}\right)
  \right\}
  \;,
\end{equation}
and
\begin{equation}\label{PA18}
 \widetilde{\Gamma}_{\bar{m}\bar{n}}^> ={\Gamma}_{\bar{m}\bar{n}}^>=
 2\pi J_0\left\{
 \left| V_{\bar{m}\bar{n}}^{(-\bar{\ell})}\right|^2\frac{1}{1-\re^{-x}}+
 \sum_{\stackrel{\ell\in{\mathbbm Z}}{\ell\neq\-bar{\ell}}} \left| V_{\bar{m}\bar{n}}^{(\ell)}\right|^2
 N\left(\omega_{\bar{m}\bar{n}}^{\ell}\right)
  \right\}
  \;.
\end{equation}
For the modified matrix  $\widetilde{\Gamma}^>$ additionally two diagonal elements will diverge for $x\to 0$.
According to
\begin{equation}\label{PA19}
 \widetilde{\Gamma}_{\bar{m}\bar{m}}^> =\Gamma_{\bar{m}\bar{m}}^>-\sum_n \Gamma_{n\bar{m}}^>
 \;,
\end{equation}
see (\ref{eq:GAT}), the diverging term of $\widetilde{\Gamma}_{\bar{m}\bar{m}}^>$ is
\begin{equation}\label{PA20}
 -2\pi\,J_0\,\left|  V_{\bar{n}\bar{m}}^{(\bar{\ell})}\right|^2 \frac{1}{\re^x-1}
 \;.
\end{equation}
Analogously, the diverging term of $\widetilde{\Gamma}_{\bar{n}\bar{n}}^>$ is
\begin{equation}\label{PA21}
 -2\pi\,J_0\,\left|  V_{\bar{m}\bar{n}}^{(-\bar{\ell})}\right|^2 \frac{1}{1-\re^{-x}}
 \;.
\end{equation}
All terms in (\ref{PA17}-\ref{PA21}) can be written as Taylor series in $x$ with the exception of the
highlighted exponential terms that possess the Laurent series
\begin{equation}\label{PA22}
  \frac{1}{\re^x-1}=\frac{1}{x}-\frac{1}{2}+\frac{x}{12}+O(x^2)
  \;,
\end{equation}
and
\begin{equation}\label{PA23}
  \frac{1}{1-\re^{-x}}=\frac{1}{x}+\frac{1}{2}+\frac{x}{12}+O(x^2)
  \;.
\end{equation}

Recall that the vector $\mathbf{ p}^>$ of NESS probabilities in the positive neighbourhood  is the (normalized) solution of
$\widetilde{\Gamma}^>\,\mathbf{ p}^>=0$ that is unique due to Theorem \ref{TUP}. After expanding $\widetilde{\Gamma}^>$ and $\mathbf{ p}^>$
into Laurent series w.~r.~t.~$x$ we will set the first three coefficients of the resulting Laurent series of
$\widetilde{\Gamma}^>\,\mathbf{ p}^>$ to zero and thus obtain the first two terms of (\ref{PA17}).
These will determine the limit of the NESS probabilities and its gradient at the phase boundary.

In order to keep the representation as simple as possible we will, without loss of generality,
assume that $\bar{n}=1$ and $\bar{m}=2$. It will suffice to give the structure of the Laurent series
of $\widetilde{\Gamma}^>$ without going into the details of how the various numbers can be expressed by the physical quantities:
\begin{equation}\label{PA24}
\widetilde{\Gamma}^> =
\left(
\begin{array}{ccc}
  -\frac{a}{x}+d+\ldots &  \frac{a}{x}+b+\ldots & {\mathbf a}_0^\top+x\,{\mathbf a}_1^\top \\
   \frac{a}{x}+c+\ldots &  -\frac{a}{x}+e+\ldots & {\mathbf c}_0^\top+x\,{\mathbf c}_1^\top \\
  {\mathbf b}_0+x\,{\mathbf b}_1 & {\mathbf d}_0+x\,{\mathbf d}_1& \boldsymbol{\gamma}_0+x\,\boldsymbol{\gamma}_1
\end{array}
\right)+O(x^2)
\;.
\end{equation}
Here we have omitted the $x$-linear terms in the upper left $2\times 2$-submatrix that are  not needed in the sequel.
The real numbers $a,b,c,d,e$ are independent of $x$, likewise the $(N-2)$-dimensional vectors
${\mathbf a}_0,\ldots, {\mathbf d}_1$ and the $(N-2)\times (N-2)$-matrices $\boldsymbol{\gamma}_0$ and $\boldsymbol{\gamma}_1$.
We stress that the repeated occurrence of the quantity
\begin{equation}\label{PA25}
  a=2\pi\,J_0\,\lim_{x\downarrow 0}\left|  V_{12}^{(\bar{\ell})}\right|^2
\end{equation}
in (\ref{PA24}) is crucial for the following considerations. The vector of NESS probabilities $\mathbf{p}^>$
will be written as
\begin{equation}\label{PA26}
  {\mathbf p}^>=\left(
 \begin{array}{c}
   p_{10}+x\,p_{11} \\
   p_{20}+x\,p_{21} \\
   {\mathbf p}_0+x\, {\mathbf p}_1
 \end{array}
 \right)+O(x^2)
 \;.
\end{equation}
Setting the coefficients of the resulting Laurent series of the various components of
$\widetilde{\Gamma}^>\,{\mathbf p}^>$ to zero yields the following results:
\begin{eqnarray}
\label{PA27a}
  x^{-1} &:& \frac{a}{x}\left( p_{20}-p_{10}\right)=0\;\Rightarrow\; p_{20}=p_{10}\equiv p\;,
  \\
  \label{PA27b}
   x^0&:&p\left({\mathbf b}_0+{\mathbf d}_0 \right)+\boldsymbol{\gamma}_0\,{\mathbf p}_0=0
   \;\Rightarrow\;
   {\mathbf p}_0=-p\,\boldsymbol{\gamma}_0^{-1}\left({\mathbf b}_0+{\mathbf d}_0 \right)\;,
   \\
    x^0 &:&
  a\,\left( \begin{array}{rr}
              1 & 1 \\
              1 & -1
            \end{array} \right)
            \left( \begin{array}{c}
                   p_{11}\\
                     p_{21}
                   \end{array} \right)=
                   \left(  \begin{array}{c}
                    -p(d+b)-{\mathbf a}_0\cdot{\mathbf p}_0\\
                     -p(c+e)-{\mathbf c}_0\cdot{\mathbf p}_0
                   \end{array} \right)\\
  \label{PA27c}
  && \Rightarrow\; p_{11}=-\frac{1}{2a}\left(p(d+b+c+e)+ {\mathbf a}_0\cdot{\mathbf p}_0
  +{\mathbf c}_0\cdot{\mathbf p}_0\right)\mbox{ and }\\
  \label{PA27d}
  &&\quad\;\; p_{21}=-\frac{1}{2a}\left(p(d+b-c-e)+ {\mathbf a}_0\cdot{\mathbf p}_0-{\mathbf c}_0\cdot{\mathbf p}_0\right)\;,\\
  \label{PA27e}
  x^1 & : & x\left( p_{11}\,{\mathbf b}_0+p\,{\mathbf b}_1+p_{21}\,{\mathbf d}_0+p\,{\mathbf d}_1+
  \boldsymbol{\gamma}_1\,{\mathbf p}_0+\boldsymbol{\gamma}_0\,{\mathbf p}_1  \right)=0\\
   \label{PA27f}
  && \Rightarrow\; {\mathbf p}_1= -\boldsymbol{\gamma}_0^{-1}\left(
  -\boldsymbol{\gamma}_1\,{\mathbf p}_0+p\left({\mathbf b}_1+{\mathbf d}_1\right)+p_{11}\,{\mathbf b}_0+p_{21}\,{\mathbf d}_0
  \right)
  \;.
\end{eqnarray}
A few remarks are in order. First, we note that the result $p_{20}=p_{10}\equiv p$ in (\ref{PA27a})
again confirms the previous statement in Assertion \ref{A1} that at least two NESS probabilities coincide
at the phase boundaries. Of course, the free parameter $p>0$ has to be chosen in such a way that the
probabilities sum up to unity.

Second, we have used in (\ref{PA27b}) and  (\ref{PA27e}) that $\boldsymbol{\gamma}_0$ is invertible.
This can be shown as follows.
Let, for $-\delta<x<\delta$, $\Gamma^\wedge(x)$ denote the matrix obtained from $\Gamma^>(x)$ by subtracting its principle part,
i.~e., the terms of the form $\pm \frac{a}{x}$, analogously for $\widetilde{\Gamma}^\wedge(x)$.
Then it can be easily shown that $\Gamma^\wedge(x)$ also satisfies the conditions of
Theorem \ref{TUP}. Hence $\widetilde{\Gamma}^\wedge(x)$ has an one-dimensional null space spanned by some
$p^\wedge>0$. This vector cannot lie in the subspace of vectors of the form $(0,0,{\mathbf p})^\top$
and the matrix $\boldsymbol{\gamma}(x)$, defined
as the restriction of $\widetilde{\Gamma}^\wedge(x)$ to this subspace, must be invertible for all $-\delta<x<\delta$.
Especially, $\boldsymbol{\gamma}_0=\boldsymbol{\gamma}(0)$ is invertible.\\

The calculations with $\widetilde{\Gamma}^<$ and $p^<$ defined in the ``negative neighbourhood"
$\boldsymbol{\mathcal P}_{\bar{n}\bar{m}\bar{\ell}}^<$
 of $\boldsymbol{\mathcal P}_{\bar{n}\bar{m}\bar{\ell}}$
given by $\omega_{\bar{n}\bar{m}}^{(\bar{\ell})} < 0$ are completely analogous and need not be given in detail.
The only difference is that for $x<0$ we have
\begin{equation}\label{PA28}
N(\omega_{12}^{(\ell)})=  \frac{1}{1-\re^x}=-\frac{1}{x}+\frac{1}{2}-\frac{x}{12}+O(x^2)
  \;,
\end{equation}
and
\begin{equation}\label{PA29}
N(\omega_{21}^{(-\ell)})=  \frac{1}{\re^{-x}-1}=-\frac{1}{x}-\frac{1}{2}-\frac{x}{12}+O(x^2)
  \;.
\end{equation}
This means that the Laurent series for $\widetilde{\Gamma}^<$ is identical with (\ref{PA24}),
with the only exception that $a$ has to be replaced by $-a$. This modification does not change the
solution for $p_{10}=p_{20}=p$ according to (\ref{PA27a}) and for ${\mathbf p}_0$ according to (\ref{PA27b}).
Hence the NESS probabilities are continuous at the phase boundaries.
In contrast, the solutions for $p_{11}$ and $p_{21}$ according to (\ref{PA27c}) and (\ref{PA27d}) will change their sign
and hence also ${\mathbf p}_1$ according to (\ref{PA27f}) will be different for the negative neighbourhood.
This means that the $x$-derivative and hence the gradient of the NESS probabilities will be discontinuous at the phase
boundaries, thereby completing the arguments in favour of Assertion \ref{A2}. \hfill$\Box$\\

\begin{acknowledgments}
I would like to thank all members of the DFG Research Unit FOR 2692, especially Martin Holthaus and J\"urgen Schnack, for stimulating and
insightful discussions and hints to relevant literature.
\end{acknowledgments}


\end{document}